\documentclass{egpubl}
\usepackage{sca2020}
\usepackage{amsmath}
\SpecialIssuePaper         % uncomment for final version of Computer Graphics Forum, special issue
\usepackage[T1]{fontenc}
\usepackage{dfadobe}  

\usepackage{cite}  % comment out for biblatex with backend=biber
\BibtexOrBiblatex
\electronicVersion
\PrintedOrElectronic
\ifpdf \usepackage[pdftex]{graphicx} \pdfcompresslevel=9
\else \usepackage[dvips]{graphicx} \fi

\usepackage{egweblnk} 
\usepackage{amsmath}
\usepackage{algorithm}
\usepackage{algpseudocode}
\usepackage{tabularx}
\usepackage{booktabs}
\usepackage{enumitem}
\usepackage{graphicx}
\usepackage{amssymb}
\usepackage{subcaption}
\usepackage{etoolbox}
\usepackage{color,soul}
\usepackage{titlesec} % to mess with section spacing
\usepackage{placeins} % for float barrier

\titlespacing*{\section}
{0pt}{1ex plus 0.5ex minus .2ex}{0pt}
\titlespacing*{\subsection}
{0pt}{1ex plus 0.5ex minus .2ex}{0pt}

\title[Probabilistic Character Motion Synthesis using a Hierarchical DLVM]{Probabilistic Character Motion Synthesis using a Hierarchical Deep Latent Variable Model}

\author[S. Ghorbani, C. Wloka, A. Etemad, M. A. Brubaker, \& N. F. Troje]
{\parbox{\textwidth}{\centering S. Ghorbani$^{1}$\orcid{0000-0002-3227-9013}, C. Wloka$^{1}$\orcid{0000-0002-0249-9306}, A. Etemad$^{2}$\orcid{0000-0001-7128-0220}, M.\,A. Brubaker$^{1}$\orcid{0000-0002-7892-9026}, and N. \,F. Troje$^{1}$\orcid{0000-0002-1533-2847}
        }
        \\
{\parbox{\textwidth}{\centering $^1$York University, Toronto, Canada\\
         $^2$Queen's University, Kingston, Canada
       }
}
}
\begin{document}
% uncomment for using teaser
\teaser{
  \includegraphics[width=0.8\linewidth]{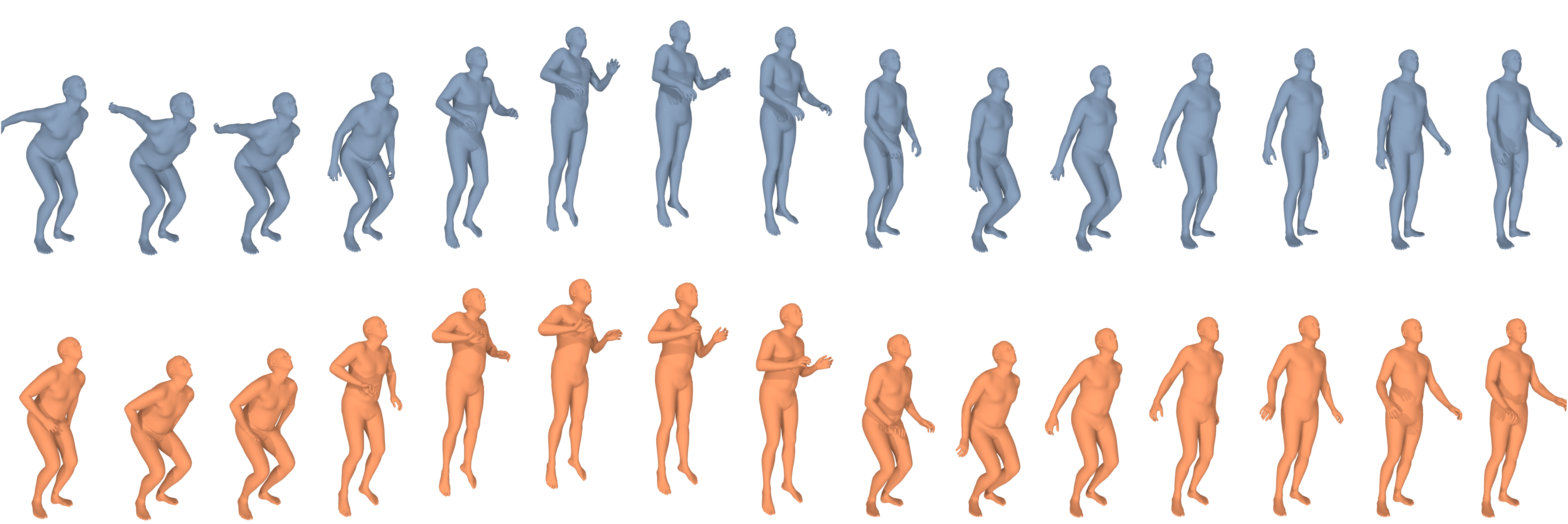}
  \centering
   \caption{Samples of a real motion sequence (blue) and synthesized motion sequence generated by our model (orange)}
 \label{fig:teaser}
}
%-------------------------------------------------------------------------
\maketitle
%-------------------------------------------------------------------------
\begin{abstract}
We present a probabilistic framework to generate character animations based on weak control signals, such that the synthesized motions are realistic while retaining the stochastic nature of human movement. The proposed architecture, which is designed as a hierarchical recurrent model, maps each sub-sequence of motions into a stochastic latent code using a variational autoencoder extended over the temporal domain. We also propose an objective function which respects the impact of each joint on the pose and compares the joint angles based on angular distance. We use two novel quantitative protocols and human qualitative assessment to demonstrate the ability of our model to generate convincing and diverse periodic and non-periodic motion sequences without the need for strong control signals.   
%-------------------------------------------------------------------------
%  ACM CCS 1998
%  (see https://www.acm.org/publications/computing-classification-system/1998)
% \begin{classification} % according to https://www.acm.org/publications/computing-classification-system/1998
% \CCScat{Computer Graphics}{I.3.3}{Picture/Image Generation}{Line and curve generation}
% \end{classification}
%-------------------------------------------------------------------------
%  ACM CCS 2012
% (see https://www.acm.org/publications/class-2012)
%The tool at \url{http://dl.acm.org/ccs.cfm} can be used to generate
% CCS codes.
%Example:

\end{abstract}
%-------------------------------------------------------------------------
\section{Introduction}\label{sec:Introduction}
An active research area in computer animation is the automatic generation of realistic character animations given a set of control parameters. This can reduce the workload of key-framing, which is a laborious and time-consuming task done by skilled animators. Recent advances in motion capture technology and deep learning methods have increased interest in data-driven and learnable frameworks for modelling human motion. Most approaches model the motion sequences as a deterministic process, meaning that for a given set of control parameters only a single, fixed sequence is generated. On the other hand, human motion is stochastic in nature - given the same intention and target the joints always travel different paths. Hence, deterministic models fail to reflect such diversity which is an essential requirement for generating convincing and realistic character animation. Another challenge in designing a motion generative model is to enforce desirable motion sequences constrained by weak control signals such as action type. This is due to the fact that deterministic models usually regress to the mean pose in the long run as no strong control signal can be provided, especially for non-periodic movements, to steer the motion and reduce the motion uncertainty over time. Most recent controllable approaches are proposed only for periodic movements with strong control signals such as trajectory characteristics \cite{holden2017, holden2016, pavllo2019}, or are limited to short-term predictions for non-periodic movements \cite{martinez2017, fragkiadaki2015, jain2016}. Our work addresses these open issues by developing a model for animation synthesis which can be modulated by weak control signals while retaining the desired stochastic characteristics of human motion across both temporal and spatial dimensions. Weak control signals are particularly useful for tasks in which large numbers of sequences are required, such as crowd simulation or providing data for other frameworks such as motion matching. In these situations requiring strong control signals such as trajectory characteristics would be unnecessarily labor intensive, whereas our framework can continuously generate novel motion clips with minimal user oversight.

Our proposed model for character animation synthesis is based on a deep recurrent neural network. We train our recurrent model on a large database of motion capture data such that it can generate novel, convincing motion samples that imitate the high-level stochastic nature of real data. This semi-supervised framework does not require any manual data preparation such as time-warping or motion clipping which minimizes the amount of manual work in the training and synthesizing processes.

Our framework is designed as a hierarchical recurrent latent variable network which models the spatiotemporal motion data with a two-level hierarchy. The hierarchical structure of the network architecture allows not only for motion sequences to be represented at multiple levels of abstraction but also for a higher level of desired variability in the generative process. The inner layer of the proposed architecture is designed as an extension of a variational recurrent neural network \cite{chung2015} which is conditioned on control signals and recursively processes high-level feature vectors (derived from motion subsequences) along with a stochastic latent variable. Defining this latent variable at a high level of abstraction enables the network to model the stochasticity observed in human movement. The inner layer is wrapped by encoder and decoder layers which encode the motion subsequences into feature vectors and decode the generated feature vectors back to motion subsequences.

We also propose a new objective function based on the geodesic distance between the ground-truth and reconstructed joint angles which has the following principal advantages: \textit{i}) The geodesic distance better represents the deviation from the desired output than $l_p$ norm losses.
\textit{ii}) The influence of different joints in the kinematic tree can be represented by assigning different weights to each joint. 
\textit{iii}) High level and semantic information are integrated into the learning process by comparing the ground-truth and reconstructed sequences in the feature space of pre-trained classifiers.

We validated the performance of our model both qualitatively, via human scoring, and quantitatively through a novel evaluation protocol based on the \emph{Inception Score (IS)} \cite{salimans2016} and \emph{Fréchet Inception Distance Score (FID)} \cite{heusel2017}. These metrics were first used for evaluation in image synthesis, and provide a measure not only of the quality of the generated output but also the diversity of output provided. Given the importance of movement variety for character animation synthesis, we have therefore adapted these metrics to provide a more complete evaluation than previously used metrics. The results show that our model effectively learns human motion dynamics and is capable of generating realistic and diverse character animation sequences coherent with control parameters, outperforming all other state-of-the-art models tested.

Our contributions can be summarized as: \textit{i}) we propose a novel hierarchical generative recurrent architecture which effectively learns human motion dynamics and generates realistic character animation sequences coherent with control parameters, \textit{ii}) we present a new objective function based on angular distances and the influence of different components in the kinematic tree which better represents network error and leads to improved learning, \textit{iii}) we provide a new benchmark and evaluation protocol for character animation synthesis to measure both the quality and variability of generated sequences. 
%-------------------------------------------------------------------------
\section{Related Work}\label{sec:Related Work}
\textbf{Traditional Data-Driven Approaches: }Data-driven approaches to character animation synthesis have been a popular area of research for nearly two decades. These approaches rely on motion capture data \cite{mahmood2019, sigal2006, ghorbani2020} which are provided as a sequence of poses represented by 3D joint locations or 3D joint angles of the skeleton at each time frame. With the advent of such motion capture datasets, many traditional approaches such as Motion Graphs \cite{arikan2002, kovar2004}, PCA-based models \cite{safonova2004}, Kernel-based models \cite{mukai2005,mukai2011}, and Hidden Markov models (HMMs) \cite{tanco2000} were proposed for the task of motion synthesis. However, these approaches fail to model the complex nonlinear dynamics of human movements, especially for long-term multi-modal motion synthesis. For instance, HMMs require a hidden state size which is exponential in the number of components and therefore suffers from having a simple hidden state. 

\textbf{Early Deep-Learning-based Approaches: }One of the earliest attempts to overcome the limitations of the above approaches, Taylor et al. \cite{taylor2007, taylor2009,taylor2011} approached human motion modelling using variations of conditional Restricted Boltzmann Machines (cRBM) as an undirected energy-based model. They modelled the temporal dependency by adding poses from previous time steps as additional inputs. More recently, the impressive results achieved by deep generative models in other areas such as image and speech synthesis has encouraged researchers to model human movement using these models. The bulk of these works make use of recurrent neural networks (RNNs) \cite{fragkiadaki2015, jain2016, martinez2017, aksan2019, Wang2019spatio, Wang2019combining, lee2018} as they have a high representational capacity in their internal state. 

\textbf{RNN-based Approaches: }A notable approach which applies the recurrence step to a learned representation was introduced by Fragkiadaki et al. \cite{fragkiadaki2015}, who proposed an Encoder-Recurrent-Decoder (ERD) architecture which encodes each pose into a feature vector where it is recursively processed through a two-layer LSTM network. During motion synthesis, the prediction is fed back to the model in the following time step which causes the accumulation of small errors at each time step (called exposure bias). To address this problem, they corrupted the input by Gaussian noise with progressively increasing standard deviation as a type of curriculum learning. While it is hard to tune the amount of noise, this strategy was used in some of the subsequent proposed approaches as well \cite{jain2016}. To tackle the problem of exposure bias, Martinez et al \cite{martinez2017} exposed the model to its own prediction during training using a seq-2-seq architecture. They also enforced the model to learn velocities instead of absolute values via residual connections to address the problem of discontinuity in the seq-2-seq models.

\textbf{Approaches Exploiting a Kinematic Tree: }Both \cite{fragkiadaki2015} and \cite{martinez2017} modelled motion without an explicit representation of the kinematics tree, but this structural information is a potentially very useful model component. By explicitly modelling spatiotemporal interactions between joints, Jain et al. \cite{jain2016} combined the explicit representational power of spatiotemporal graphs with the implicit sequential learning of RNNs. Aksan et al. \cite{aksan2019} demonstrated an alternative method of incorporating structural information, proposing a Structured Prediction Layer (SPL) where the prediction of each joint at time $t$ is conditioned on the joint's previous state and the current state of the parent joint. Therefore, at each time step, the joint angles are predicted starting from the root to the leaf nodes in the kinematic tree. They integrated the proposed layer in a variety of baseline architectures and showed improvements for the task of motion prediction.

\textbf{Approaches based on Strong Control Signals: }Although RNN-based methods overall show impressive results in short-term motion prediction, they fail in long-term generation due to their deterministic state assumption which fails to capture the intrinsic variability in human motion which compounds through time. Additionally, deterministic models assume a single future output, which causes them to converge over the long-term to a mean pose (referred to as \emph{mean collapse}). Some of the proposed methods addressed this problem by adding additional information to the model to disambiguate the generative process. Holden et al \cite{holden2017, holden2016} proposed providing foot contact and phase information as strong control parameters for locomotion movements to decrease the model uncertainty. Pavllo et al \cite{pavllo2019} augmented the generative network with a pre-trained \emph{pace network} which provides the foot-step frequency, local speed, and facing direction given the trajectory. Martinez et al \cite{martinez2017} showed concatenating weak control parameters such as action type to the input sequence alleviates the mean collapse problem to some extent. 

\textbf{Probabilistic Approaches: }Another way of avoiding converging to the mean pose is to model the intrinsic uncertainty of motion using probabilistic schemes. Many approaches were proposed based on Gaussian Process Latent Variable Models (GPLVM) \cite{levine2012,wang2006}. However, these models are limited due to their high memory cost for large data. More recently, adversarial learning has also been investigated for non-deterministic human motion modelling. Barsoum et al \cite{barsoum2018} proposed a probabilistic motion prediction approach via GANs. Their model architecture is designed based on a seq-2-seq model and predicts multiple possible sequences from the same input.  However, GANs are oftentimes hard to train, and their method was not designed to be steered by control signals. Using an alternative method also originally derived from image synthesis, Henter et al. \cite{henter2019} proposed an autoregressive model based on normalizing flows (NFs) \cite{rezende2015,dinh2014}. They extended a variant of NFs, \emph{GLOW} \cite{kingma2018}, to bipedal and quadrupedal motion sequences. However, their model needs strong trajectory control signals and is limited to locomotion synthesis.  Similar to our proposed model, Habibie et al. \cite{habibie2017} use a variational autoencoder (VAE) to model the spatial relationships. However, they extended their model to operate in an autoregressive fashion by setting the cell state of the LSTM components equal to the corresponding latent variable at each time step during training. While this approach successfully couples the LSTM representation to the posterior distribution, by collapsing the latent variable and internal state to one variable it limits the representational power of the model's internal state. Additionally, the balance their architecture strikes between control signals and previous cell state during generation limits model performance for non-periodic complex movements. We attempt to mitigate these drawbacks in our proposed model by formulating the internal state and latent code in two separate channels and conditioning the latent code to the previous internal state to model the temporal dependencies during test time. Recently, \cite{Ling2020} et al. proposed an interesting model based on VAEs where the motion is controlled by setting the latent code as the output of a deep reinforcement learning module. Unlike our method, they modelled the motion by a Markovian assumption, meaning that each pose only depends on the previous pose and the autoregressive model is memoryless. They also modelled the VAE decoder as a Mixture of Experts (MoE) network.

\textbf{Mixture-of-Experts Approaches: }
Another strategy exploited in \cite{Starke2019, Starke2020, Ling2020} to address the problem of mean collapse in multi-modal motion data is to use a Mixture-of-Experts (MoE) network where each expert is responsible for one mode in the training data. Though effective at mitigating mean collapse, the number of parameters in these networks increases with the number of experts.

\textbf{Style Transfer Approaches: }Motivated by recent advances in style transfer in images and videos, style transfer techniques were exploited to transfer the style of one animation clip to another \cite{smith2019, aberman2020}. While this method generates natural motion sequences with the desired style, these approaches cannot be directed by other control signals.
%-------------------------------------------------------------------------
%-------------------------------------------------------------------------
\section{System Overview}\label{sec:System Overview}
\begin{figure}[t]
  \centering
  \includegraphics[width=0.9\linewidth]{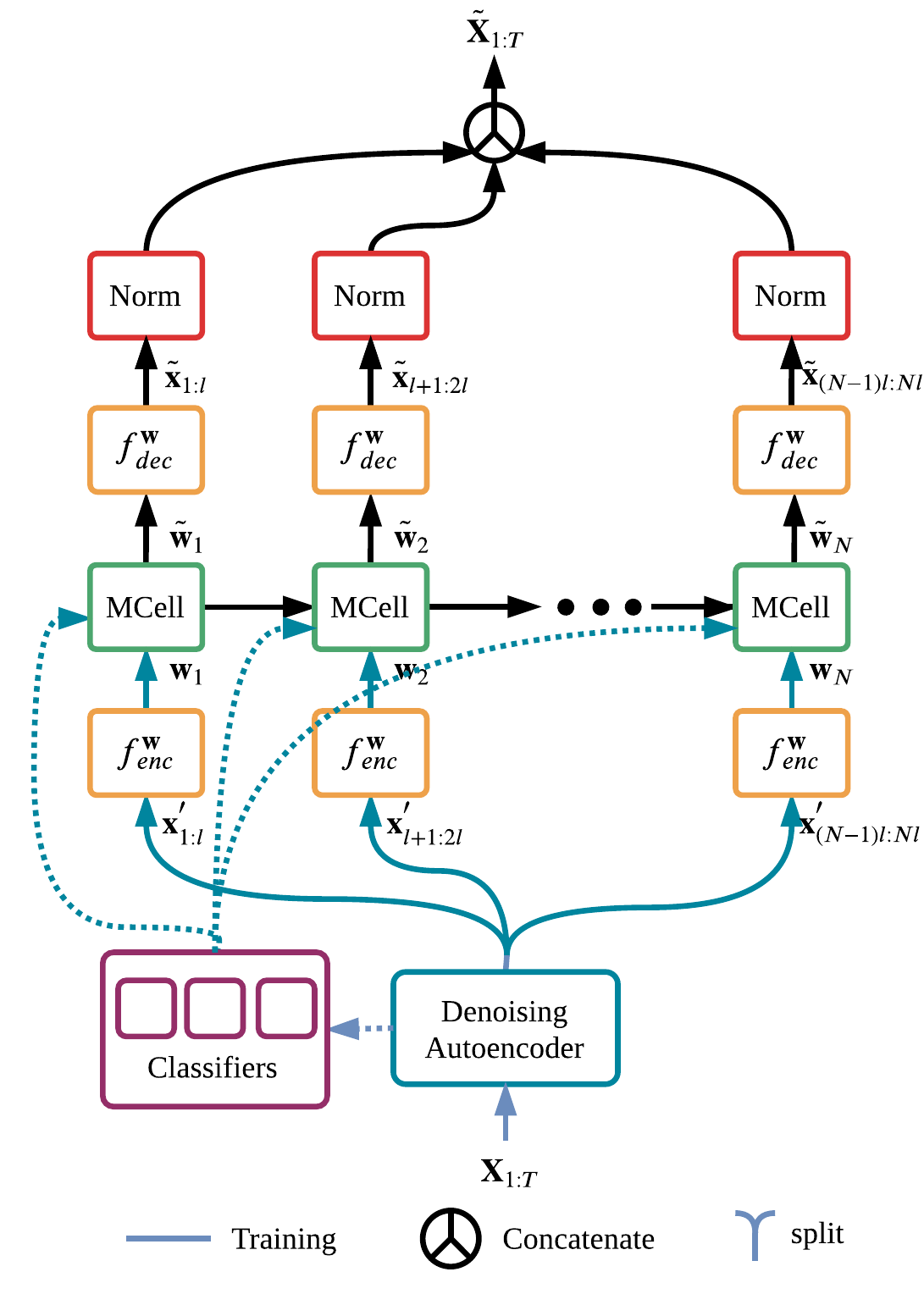}
  \caption{\label{fig:main_model}
            % #2
           An overview of our recurrent model. During training, denoised frames form temporal windows of equal size where each window $\mathbf{x}^{'}_{(i-1)l+1:il}$ is projected into a high level feature vector $\mathbf{w}_{i}$ (called a Motion Word) via $f^w_{enc}$. Motion Cells operate on Motion Words in a latent space to combine information from the preceding sequence with stochastic variability to output the next step in the sequence. This output is converted to a joint angle representation via $f^w_{dec}$. A set of classifiers provide control signals for unlabelled input and normalization ensures that the representations fall within valid ranges.}
           \vspace{-5mm}
\end{figure}
%The proposed model aims to synthesize realistic character animation controlled by attribute vectors while retaining a high-level of natural variability and stochasticity.  To this end, we propose a controllable hierarchical recurrent latent variable network which models this variability by use of latent variables. 
A visual diagram of our model architecture is given in Fig \ref{fig:main_model}. We provide a framework which encapsulates both the hierarchical and the stochastic nature of human motion within a deep hierarchical recurrent architecture. Our model generates motion sequences via a two-level hierarchy. In particular, we model the human motion as a sequence of high-level feature vectors called \emph{Motion Words} where each Motion 
Word, $\mathbf{w}_i$, is computed as a function of a sub-sequence of poses. The recurrent processing of motion sequences is thereby applied at word-level rather than at pose-level.

We leverage an extension of a variational recurrent neural network~\cite{chung2015} which contains a variational autoencoder at each time-step conditioned on the control signals. We call the recurrent processing unit a \emph{Motion Cell} (green blocks in Fig \ref{fig:main_model}) which attaches a stochastic latent variable to the observed Motion Words at each time-step. Stochasticity at the Word level enables variability to be represented at a higher level of abstraction (see section \ref{sec:Hierarchical Probabilistic Recurrent Network} for details), thereby producing more internally consistent motion sequences. The mapping between Motion Words and the corresponding sub-sequence of poses is performed by $f_{enc}$ and $f_{dec}$ (yellow blocks in Fig \ref{fig:main_model}). At each time step, we condition the Motion Cell on the control signals to modulate the motion characteristics and decrease uncertainty due to the multi-modality nature of the motion generation process. In general, any motion-related attribute, static or dynamic, such as style, action type, or motion trajectory could be used as a control signal. However, in this work, we used a set of holistic attributes consisting of action type and gender.

During training, we integrated individual pre-trained classifiers to the model for each attribute type to infer attributes from the unlabelled input sequence and also to provide additional higher-level learning signals to the objective function. This constrains the model to generate animations which fulfil the semantics defined by the attribute codes (see section \ref{sec:Attributes Classifiers} for details). 

Our model uses a joint angle representation to define each pose. We tested the model with three different joint angle representations: axis-angle vectors, quaternions, and rotation matrices. To ensure that the model produces valid rotation for each joint, the estimated rotations in the output of the hierarchical recurrent neural network are normalized into valid rotations (red blocks in Fig \ref{fig:main_model}). Regardless of the specific joint angle representation used, our model otherwise operates identically from one representation to the next. To have valid rotations represented by quaternions, the magnitude of the quaternions should be one. Therefore, we simply divided each quaternion by its magnitude. When instead using rotation matrices to represent joint angles, we applied the Gram–Schmidt orthonormalisation process on the output matrices. No normalization step was applied to the axis-angle vector representation.
%-------------------------------------------------------------------------
\subsection{Data Preprocessing}\label{sec:Data Preprocessing}

The local joint angle representation is augmented with processed root joint information which encodes the global transformation while keeping the final representation invariant to ground-plane (x-y) translation and rotation about the gravity direction. The augmented data includes forward direction velocity, sideways direction velocity, global root height, angular velocity around the gravitational axis, and the pitch and roll relative to the direction where the subject is facing. During motion synthesis, global translation and orientation can be recovered by integrating velocities over time while we assume the initial facing direction is in the direction of the x-axis in the global coordinate system. The final pose representation consists of a $D_p = 21 \times k + 6$ dimensional vector, where $k$ is $3, 4$ or $9$ for axis-angle vectors, quaternions,  and rotation matrices, respectively. We sub-sampled the motion sequences into 30 frames per second (they were originally recorded in 120 frames per second) and used all four offsets for training. 

Our model can be trained by variable-length sequences of inputs. However, to speed up the training process by parallel computing we set the size of input sequences to a fixed size by clipping the longer sequences and padding zeros to the shorter ones. In our work, we set the length of input sequences to $200$ frames (around $6.6$ seconds). For synthesis the length of a generated sequence does not have to be equal to the length of the training sequences, rather our model can generate sequences with arbitrary length.

Before feeding to the main model, we apply a pre-trained denoising network to the training sequences to correct possible errors in the training data such as high-frequency noise resulting from marker occlusions or mislabelled markers in the motion capture process. The denoising network is implemented as a one-dimensional convolutional denoising autoencoder pretrained on a different subset of data than the one we used for training the main model. Details of the denoising network structure and its training process are given in Section \ref{sec:Training Process}.
%-------------------------------------------------------------------------
\subsection{Hierarchical Probabilistic Recurrent Network}\label{sec:Hierarchical Probabilistic Recurrent Network}
Our hierarchical recurrent network models a motion sequence of length $T$ with a two-level hierarchy. In the pose-level, we have a sequence of poses and in the word-level, we have a sequence of Motion Words. Each Motion Word, $\mathbf{w}_n \in \mathbb{R}^{D_w}$, summarizes a sub-sequence of poses using an encoding function
\begin{equation}
    \mathbf{w}_n = f_{\text{enc}}^{\mathbf{w}}(\mathbf{X}_{nl:(n+1)l}),
\end{equation}
where $\mathbf{X}_{1:T}$ is the sequence of poses ($\mathbf{X}_{t}\in \mathbb{R}^{D_p}$),  $f_{\text{enc}}^{\mathbf{w}}$ is a non-linear complex encoder such as a fully-connected neural network, and $l$ is the length of each sub-sequence.
We define the sequence of Motion Words as an autoregressive model as follows:
\begin{equation}
    p\left(\mathbf{w}_{1}, \ldots, \mathbf{x}_{N}\right)=p\left(\mathbf{w}_{1}\right)\prod_{t=2}^{N} p\left(\mathbf{w}_{n} | \mathbf{w}_{< n}\right) ,
\end{equation}
where $N=\lfloor{T}/{l}\rfloor$ is the number of Motion Words in the sequence. $l$ is considered as a hyperparameter where the best results were achieved for $l = 3$ . The dimension of the Motion Word $D_w$ was set to $32 * 3 = 96$ which is equal to the approximate degrees of freedom in each pose (\cite{Ling2020,peng2018}) times the length of each subsequence. To model the recursive structure of Motion Words we used a variational recurrent neural network \cite{chung2015} extended to condition on the control parameters. The proposed recursive model can be formulated as a recurrent neural network built upon a probabilistic recurrent cell which is structured as a conditional VAE at each time-step. We call these recurrent cells \emph{Motion Cells} (green \emph{MCell} blocks in Fig \ref{fig:main_model}, see section \ref{sec:Motion Cell} for more details). The combination of a hierarchical structure and probabilistic recurrence allows the model to define a stochastic latent variable at the word-level. Hence, the stochastic behaviour of the generation process is modelled at a deep level using highly abstracted features, allowing variation to be more easily sampled in an internally consistent manner from the learned feature space. We use another fully connected layer to convert the generated Motion Words back to the sub-sequence of poses
\begin{equation}
    \tilde{\mathbf{X}}_{nl:(n+1)l} = f_{\text{dec}}^{\mathbf{w}}(\tilde{\mathbf{w}}_n),
\end{equation}
where $\tilde{\mathbf{w}}_n$ is the output of the Motion Cell at time-step $n$ (also called the reconstructed Motion Word) and $\tilde{\mathbf{X}}_{nl:(n+1)l}$ is the corresponding reconstructed sub-sequence. The details of the Motion Word encoder ($f_{\text{dec}}^{\mathbf{w}}$) and decoder ($f_{\text{dec}}^{\mathbf{w}}$) are given in Table \ref{tab:architecture}, and the next section describes the internal structure of a Motion Cell.
%-------------------------------------------------------------------------
\subsection{Motion Cell}\label{sec:Motion Cell}
\begin{figure}[t]
  \centering
  \includegraphics[width=1\linewidth]{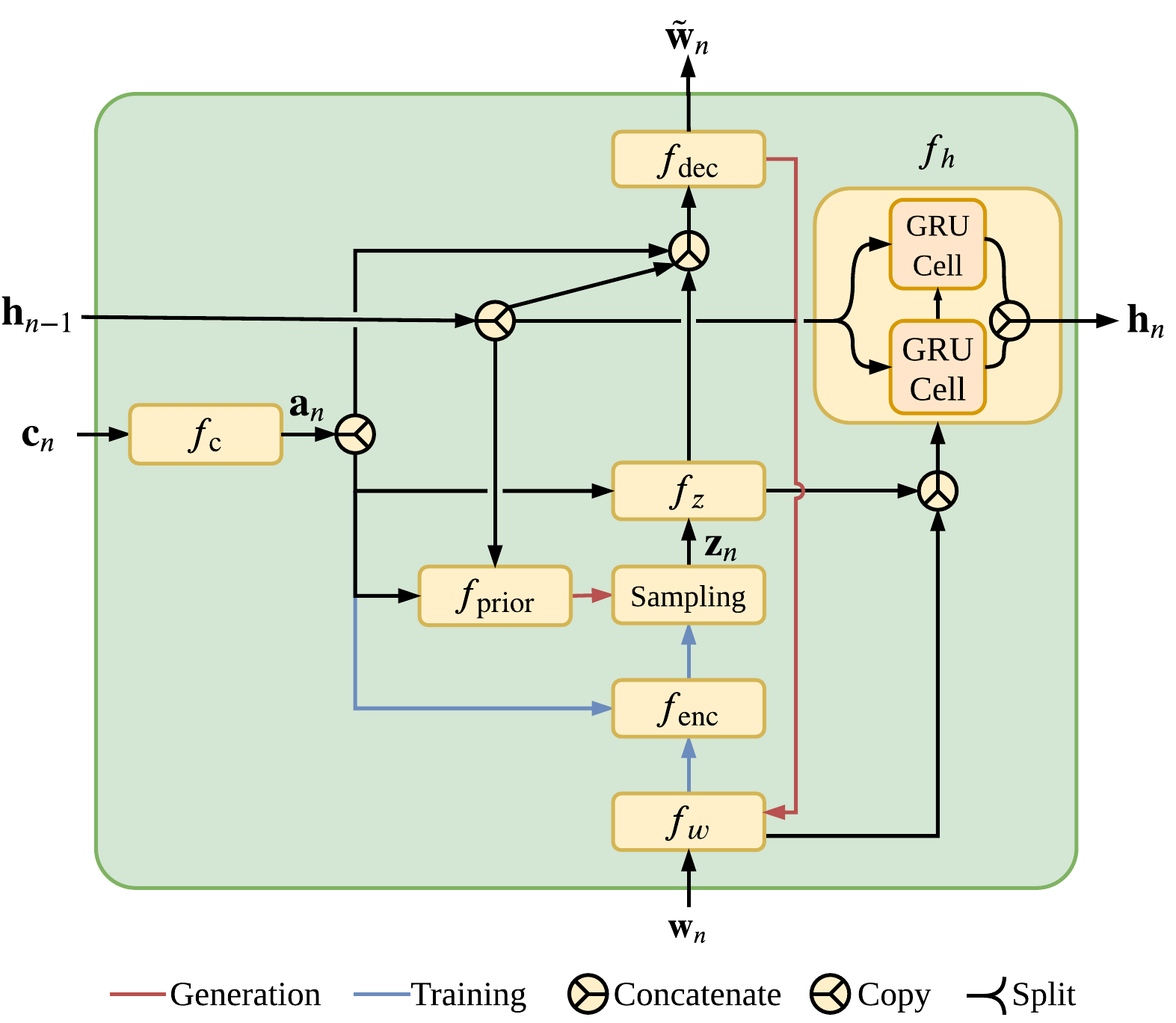}
  \caption{\label{fig:motion_cell}
           Internal structure of Motion Cell. A Motion Cell can be viewed as a recurrent unit conditioned on control signals.}
           \vspace{-5mm}
\end{figure}
Our recurrent model is constructed by a probabilistic recurrent unit called a Motion Cell. The design of a Motion Cell is based on an entangled conditional VAE and a transition block. The VAE models the spatial dependencies and is additionally conditioned on control parameters and previous information. The transition block models the temporal dependencies and is a function of not only the input variable and previous internal state, but also the current latent variable. By conditioning both spatial and temporal paths on the latent variable, we introduce variability across both dimensions. The structure of a Motion Cell is illustrated in Fig \ref{fig:motion_cell}. In the following, we describe in more detail how Motion Cells operate during training and generation phases.

\subsubsection{Training Phase} \label{sec:Training Phase}
\begin{algorithm}[t]
\caption{This algorithm represents the FORWARD process of a Motion Cell for a single time-step during \textbf{training}. It takes as input a motion word $\mathbf{w}_{n}$, control signal $\mathbf{c}_{n}$, and previous internal state $\mathbf{h}_{n-1}$, and outputs a motion word $\tilde{\mathbf{w}}_{n}$ and updates the internal state to $\mathbf{h}_{n}$}
\label{alg:1}
\begin{algorithmic}
\Function{Forward}{$\mathbf{w}_{n}$, $\mathbf{c}_{n}$, $\mathbf{h}_{n-1}$}
    \State $\mathbf{a}_{n} = f_{\text{c}}(\mathbf{c}_{n})$
    \State Compute Posterior distribution
    \State$\boldsymbol{\mu}_{q, n}=f_{\text{enc}}^{\boldsymbol{ \mu}}\left(f_{\mathbf{w}}\left(\mathbf{w}_{n}\right), \mathbf{h}_{n-1}, \mathbf{a}_n\right)$
    \State$\boldsymbol{\sigma}_{q, n}=f_{\text {enc}}^{\boldsymbol{ \sigma}}\left(f_{\mathbf{w}}\left(\mathbf{w}_{n}\right), \mathbf{h}_{n-1}, \mathbf{a}_n\right)$
    \State Sample latent variable from Posterior distribution 
    \State (using reparameterization trick)
    \State $\mathbf{z}_{n}\sim\mathcal{N}(\mathbf{z}_{n} ; \boldsymbol{\mu}_{q, n}, \operatorname{diag}\boldsymbol{\sigma}_{q, n}^{2})$
    \State Compute Prior distribution
    \State $\boldsymbol{\mu}_{p, n}=f_{\text {prior }}^{\mu}\left(\mathbf{h}_{n-1}, \mathbf{a}_{n}\right)$
    \State $\boldsymbol{\sigma}_{p, n}=f_{\text {prior }}^{\sigma}\left(\mathbf{h}_{n-1}, \mathbf{a}_{n}\right)$
    \State Update internal state 
    \State $\mathbf{h}_{n}=f_{\mathbf{h}}\left(f_{\mathbf{w}}\left(\mathbf{w}_{n}\right), f_{\mathbf{z}}\left(\mathbf{z}_{n}\right), \mathbf{h}_{n-1}\right)$
    \State Compute cell output
    \State $\tilde{\mathbf{w}}_n=f_{\mathrm{dec}}\left(f_{\mathbf{z}}\left(\mathbf{z}_{n}\right), \mathbf{h}_{n-1}, \mathbf{a}_{n}\right)$
    \State \Return ($\tilde{\mathbf{w}}_n$, $\mathbf{h}_{n}$, $\boldsymbol{\mu}_{q, n}$, $\boldsymbol{\sigma}_{q, n}$, $\boldsymbol{\mu}_{p, n}$, $\boldsymbol{\sigma}_{p, n}$)
\EndFunction
\end{algorithmic}
\end{algorithm}
Algorithm \ref{alg:1} provides the processing steps of a Motion Cell for a single time-step during training (the FORWARD function). Unlike a standard VAE, the posterior is not only conditioned on the input (observation) but also on the previous internal state and control parameters.  A computationally inexpensive and common choice for the latent code distribution is a factorized Gaussian distribution
\begin{equation}
    \begin{aligned}
        q(\mathbf{z}_{n}|\mathbf{w}_{n}, \mathbf{h}_{n-1}, \mathbf{a}_{n}) &= q(\mathbf{z}_{n}|\mathbf{w}_{\leq n},\mathbf{z}_{<n}, \mathbf{a}_{\leq n}) \\
        &= \mathcal{N}(\mathbf{z}_{n} ; \boldsymbol{\mu}_{q, n}, \operatorname{diag}(\boldsymbol{\sigma}_{q, n}^{2})),
    \end{aligned}
\end{equation}
where $\mathbf{z}_{n} \in \mathbb{R}^{D_z}$ is the latent variable, $\mathbf{h}_{n} \in \mathbb{R}^{D_h}$ is the internal state of the Motion Cell which summarizes all the past information up to step $n$, and $\mathbf{a}_{n} = f_{\text{c}}(\mathbf{c}_n)\in \mathbb{R}^{D_a}$ is the feature vector extracted from control signals. In our model we only used weak attributes such as action type or style, either included as a component of sample labelling or inferred by integrated classifiers if the sample is unlabelled. However, the same methods can be straightforwardly extended to include other attributes, including dynamic parameters of the motion such as locomotion trajectory. We set $D_z=96$, $D_h=1024$, and $D_a=8$.

%$\mathcal{N}(\mathbf{z}_{n} ; \boldsymbol{\mu}_{q, n}, \operatorname{diag}(\boldsymbol{\sigma}_{q, n}^{2}))$ represents a univariate Gaussian distribution parameterized by a mean vector $\boldsymbol{\mu}_{q, n}$ and a diagonal covariance matrix of $diag(\boldsymbol{\sigma}_{q, n})$. 
The mean, $\boldsymbol{\mu}_{q, n}$, and covariance parameters,  $diag(\boldsymbol{\sigma}_{q, n})$, are computed as:
\begin{equation} 
    \begin{aligned}
        \boldsymbol{\mu}_{q, n}&=f_{\text{enc}}^{\boldsymbol{ \mu}}\left(f_{\mathbf{w}}\left(\mathbf{w}_{n}\right), \mathbf{h}_{n-1}, \mathbf{a}_n\right),\\
        \boldsymbol{\sigma}_{q, n}&=f_{\text {enc}}^{\boldsymbol{ \sigma}}\left(f_{\mathbf{w}}\left(\mathbf{w}_{n}\right), \mathbf{h}_{n-1}, \mathbf{a}_n\right),
    \end{aligned}
\end{equation}
where $f_{\text{enc}}^{\boldsymbol{ \mu}}$ and $f_{\text{enc}}^{\boldsymbol{ \sigma}}$ are non-linear complex functions such as multilayer perceptrons (MLP). $f_{\mathbf{w}}$ is also implemented as an MLP for extracting Motion Word features, which is an essential requirement for learning complex motions. During training the latent variable is sampled from the posterior distribution using the reparameterization trick \cite{Kingma2013}.
\begin{equation}
\mathbf{z}_{n} \sim \mathcal{N}\left(\mathbf{z}_{n} ; \mu_{q, n}, \operatorname{diag} \sigma_{q, n}^{2}\right)
\end{equation}
Similar to the posterior distribution, the prior distribution is also conditioned on the previous internal state and attribute vectors
\begin{equation}\label{eq:prior}
    \begin{aligned}
        p(\mathbf{z}_{n}|\mathbf{h}_{n-1}, \mathbf{a}_{n}) &= p(\mathbf{z}_{n}|\mathbf{w}_{< n},\mathbf{z}_{<n}, \mathbf{a}_{\leq n})\\
        &=\mathcal{N}(\mathbf{z}_{n} ; \boldsymbol{\mu}_{p, n}, \operatorname{diag}\boldsymbol{\sigma}_{p, n}^{2}),
    \end{aligned}
\end{equation}
where:
\begin{equation}
    \begin{aligned}
        \boldsymbol{\mu}_{p, n}&=f_{\text{prior}}^{\boldsymbol{ \mu}}\left(\mathbf{h}_{n-1}, \mathbf{a}_n\right),\\
        \boldsymbol{\sigma}_{p, n}&=f_{\text {prior }}^{\boldsymbol{ \sigma}}\left(\mathbf{h}_{n-1}, \mathbf{a}_n\right).
    \end{aligned}
\end{equation}
where $f_{\text{prior}}^{\boldsymbol{ \mu}}$ and $f_{\text{prior}}^{\boldsymbol{ \sigma}}$ are implemented as MLPs. Conditioning the prior and posterior distributions on past information increases the temporal representational power of the model. Additionally, conditioning them on control parameters helps the model find distinct modes within the latent space.

In contrast to standard RNNs in which the output distribution is only conditioned on the previous internal state, the output distribution in the Motion Cell is also conditioned on the latent variable and control signals.
\begin{equation}
    p(\mathbf{w}_{n}|\mathbf{z}_{n}, \mathbf{h}_{n-1}, \mathbf{a}_{n}) = p(\mathbf{w}_{n}|\mathbf{w}_{< n},\mathbf{z}_{\leq n}, \mathbf{a}_{\leq n}).
\end{equation}
In this work, we formulate the VAE decoder function deterministically, such that the reconstructed output, $\tilde{\mathbf{w}}_n$, is computed by an MLP:
\begin{equation}\label{eq:mcell_output}
        \tilde{\mathbf{w}}_n=f_{\text{dec}}\left(f_{\mathbf{z}}\left(\mathbf{z}_{n}\right), \mathbf{h}_{n-1}, \mathbf{a}_n\right),
\end{equation}
where $f_{\mathbf{z}}$ is a feature extraction MLP applied on the latent variable.

The internal state of the Motion Cell is updated by a transition function given the current input, previous internal state, and current latent variable:
\begin{equation}\label{eq:transition_function}
    \mathbf{h}_{n}=f_{\mathbf{h}}\left(f_{\mathbf{w}}\left(\mathbf{w}_{n}\right), f_{\mathbf{z}}\left(\mathbf{z}_{n}\right), \mathbf{h}_{n-1}\right).
\end{equation}
Conditioning the internal state on the latent variable makes the temporal transition probabilistic and also helps the model address the mean collapse problem. Similar to \cite{pavllo2019}, we used two stacked Gated Recurrent Units (GRU) with an internal state of size $512$ for the transition function where the Motion Cell internal state is formed by concatenating the internal state of the two GRU cells. All of the components of the Motion Cell are learned by optimizing the objective function explained in section \ref{sec:Objective Function}.

%----------------------------------------------------------------------------------
\subsubsection{Generation Phase}\label{Generation Phase}
\begin{algorithm}[t]
\caption{This algorithm represents the SAMPLE process of a Motion Cell for a single time-step during \textbf{generation}. It takes control signal $\mathbf{c}_{n}$ and previous internal state $\mathbf{h}_{n-1}$, and generates motion word $\tilde{\mathbf{w}}_{n}$ and current internal state $\mathbf{h}_{n}$}
\label{alg:2}
\begin{algorithmic}
\Function{Sample}{$\mathbf{c}_{n}$, $\mathbf{h}_{n-1}$}
    \State $\mathbf{a}_{n} = f_{\text{c}}(\mathbf{c}_{n})$
    \State $\boldsymbol{\mu}_{p, n}=f_{\text {prior }}^{\mu}\left(\mathbf{h}_{n-1}, \mathbf{a}_{n}\right)$
    \State $\boldsymbol{\sigma}_{p, n}=f_{\text {prior }}^{\sigma}\left(\mathbf{h}_{n-1}, \mathbf{a}_{n}\right)$
    \State Sample latent variable from Prior distribution 
    \State (using reparameterization trick)
    \State $\mathbf{z}_{n}\sim\mathcal{N}(\mathbf{z}_{n} ; \boldsymbol{\mu}_{p, n}, \operatorname{diag}\boldsymbol{\sigma}_{p, n}^{2})$
    \State Compute cell output
    \State $\tilde{\mathbf{w}}_n=f_{\mathrm{dec}}\left(f_{\mathbf{z}}\left(\mathbf{z}_{n}\right), \mathbf{h}_{n-1}, \mathbf{a}_{n}\right)$
    \State Update internal state 
    \State $\mathbf{h}_{n}=f_{\mathbf{h}}\left(f_{\mathbf{w}}\left(\tilde{\mathbf{w}}_{n}\right), f_{\mathbf{z}}\left(\mathbf{z}_{n}\right), \mathbf{h}_{n-1}\right)$
    \State \Return ($\tilde{\mathbf{w}}_n$, $\mathbf{h}_{n}$)
\EndFunction
\end{algorithmic}
\end{algorithm}
Algorithm \ref{alg:2} provides the processing steps for a single time-step of a Motion Cell during motion synthesis (the SAMPLE function). At each time step during generation the latent variable is sampled from a prior distribution, computed in the same manner as the posterior distribution sampling done in the training phase (Eq. \ref{eq:prior})
\begin{equation}
\mathbf{z}_{n} \sim \mathcal{N}\left(\mathbf{z}_{n} ; \boldsymbol{\mu}_{p, n}, \operatorname{diag} \sigma_{p, n}^{2}\right).
\end{equation}
The latent variable is then used with the previous internal state and control signals to generate the reconstructed Motion Word $\tilde{\mathbf{w}}_n$ (Eq. \ref{eq:mcell_output}). Finally, the internal state is updated using the previous internal state, current latent vector, and the reconstructed Motion Word.
\begin{equation}
\mathbf{h}_{n}=f_{\mathbf{h}}\left(f_{\mathbf{w}}\left(\tilde{\mathbf{w}}_{n}\right), f_{\mathbf{z}}\left(\mathbf{z}_{n}\right), \mathbf{h}_{n-1}\right).
\end{equation}
%-------------------------------------------------------------------------
\subsection{Attributes Classifiers}\label{sec:Attributes Classifiers}
For each attribute type, we integrate a separate pre-trained classifier into the generative model. Integrating classifiers into the hierarchical probabilistic recurrent network serves three purposes. First, they provide control parameters to the generative model for unlabelled data, allowing our system to operate in a semi-supervised manner (dashed arrows in Fig \ref{fig:main_model}). Second, during training, the classifiers provide additional high-level signals (both from their intermediate layers as well as the output class inferred by the classifier) to the objective function. This constrains the generative model to generate motions semantically coherent with the motion attributes (see Section \ref{sec:Classifiers Loss}). Third, the classifiers can be used for the evaluation of our generative model (see Section  \ref{sec:Experiments and Evaluation}).

We implemented all the classifiers using one-dimensional convolutional neural networks and trained them on $50\%$ of the training data. We observed that this amount of training data is sufficient to label the rest of the data with a high accuracy. Further details of classifier implementation are given in Section \ref{sec:Implementation and Training}.
%-------------------------------------------------
\subsection{Objective Function}\label{sec:Objective Function}
We formulate model training as an optimization problem by minimizing the objective function
\begin{equation}\label{eq:objective_function}
      \mathcal{L}= \mathcal{L}_{RVAE}+\lambda_{CL} \mathcal{L}_{CL}+ \lambda_{Ang} \mathcal{L}_{Ang}, 
\end{equation}
where $\mathcal{L}_{RVAE}$ is the recurrent VAE loss equal to the sum of the negative step-wise variational lower bound over the whole sequence. We define a new hierarchical geodesic loss for reconstruction part of $\mathcal{L}_{RVAE}$ which is more accurate than the $l_p$ norm loss and takes into account the relative impact of each joint in the kinematic tree on the final loss. $\mathcal{L}_{CL}$ is the complementary loss provided by the classifiers, found by evaluating the ground-truth and reconstructed samples in the intermediate and last layer of each classifier. $\mathcal{L}_{Ang}$ is the sum of constraints encouraging the model to produce valid joint representation. We will describe each term in more detail below.

%-------------------------------------------------------------------------
\subsubsection{RVAE Objective}\label{sec:VAE Objective}
The first term in our objective function, $\mathcal{L}_{RVAE}$, is defined as a variational autoencoder objective summed over all sequence steps as follows
\begin{equation}\label{eq:vae_loss}
    \begin{aligned}
        \mathcal{L}_{RVAE} &= \mathbf{E}_{q\left(\mathbf{z}_{\leq N} | \mathbf{w}_{\leq N}, \mathbf{a}_{\leq N} \right)}\biggr[\sum_{n=1}^{N=T/l}-\log p\left(\mathbf{w}_{n} | \mathbf{z}_{\leq n}, \mathbf{w}_{<n},  \mathbf{a}_{\leq n}\right)\\ &+ \lambda_{KL}\mathrm{KL}(q(\mathbf{z}_{n} | \mathbf{w}_{\leq n}, \mathbf{z}_{<n},  \mathbf{a}_{\leq n}) \| p(\mathbf{z}_{n} | \mathbf{w}_{<n}, \mathbf{z}_{<n},  \mathbf{a}_{\leq n}))\biggr].\\
        &=\mathcal{L}_{rec} + \lambda_{KL}\mathcal{L}_{KL}
    \end{aligned}
\end{equation}
The first term in the above loss is the expected log-likelihood or reconstruction loss which is usually defined as the distance between observations and the reconstructed values. We define our reconstruction term as a custom loss over joint angles rather than Motion Words to simultaneously train the Motion Word encoder $f_{\text{enc}}^{\mathbf{w}}$ and decoder $f_{\text{dec}}^{\mathbf{w}}$. The second term in the summation is the KL-divergence between the posterior and the prior at time-step $n$ weighted by $\lambda_{KL}$. To prevent optimization process from getting stuck in an undesirable stable equilibrium we used an annealing scheduler for $\lambda_{KL}$ where the optimization is performed for a few epochs with $\lambda_{KL} = 0$ (warm-up phase), then $\lambda_{KL}$ is slowly increased from $0$ to $1$ (annealing phase), and then for the last few epochs we set $\lambda_{KL} = 1$ (cool-down phase) \cite{bowman2015}. In the following we describe how the reconstruction loss is formulated.

\textbf{Geodesic Distance of Joint Angles: }Assuming a deterministic prediction in joint angles, the first term can be defined as a reconstruction loss. Often, metrics in the Euclidean space such as $l_1$ and $l_2$ norms are used as the reconstruction loss. However, these metrics do not represent the geodesic distance of two rotations which confuses the training process especially for large angular distances and at the beginning of the optimization process. 
To address the above-mentioned problems in Euclidean distances, we define more relevant distance functions which respect the intrinsic structure of 3D rotations both for quaternions and rotation matrices. The angular distance between two unit quaternions $\mathbf{q}$ and $\tilde{\mathbf{q}}$ is defined as 
\begin{equation}\label{eq_temp}
    d(\mathbf{q},\tilde{\mathbf{q}}) = \mathbf{q}\tilde{\mathbf{q}}^{-1} = 2 \arccos{(\mathbf{q}\cdot\tilde{\mathbf{q}})}.
\end{equation}
Since the quaternions double-cover the space of rotations meaning that quaternions $\mathbf{q}$ and $-\mathbf{q}$
represent the same rotation we can take into account this ambiguity by modifying the above function as 
\begin{equation}\label{eq:quat_loss_1}
    d_{1}(\mathbf{q},\tilde{\mathbf{q}}) =2 \arccos{(|\mathbf{q}\cdot\tilde{\mathbf{q}}|)}\\
\end{equation}
Since $\arccos$ is a monotonically decreasing function we can define an approximate but computationally less expensive distance as
\begin{equation}\label{eq:quat_loss_2}
   d_{2}(\mathbf{q},\tilde{\mathbf{q}}) = 1 - |\mathbf{q}\cdot\tilde{\mathbf{q}}|.
\end{equation}
which only needs $4$ multiplication and $1$ comparison for each pair of quaternions \cite{huynh2009}. It can be proven that the square of the $l_2$ norm of two unit quaternions is equivalent to Eq.\ref{eq:quat_loss_2} for small angular distances
\begin{equation}
    \begin{aligned}
        \|\mathbf{q}-\tilde{\mathbf{q}}\|^{2} &= \|\mathbf{q}\|^{2} + \|\tilde{\mathbf{q}}\|^{2} - 2 (\mathbf{q}\cdot\tilde{\mathbf{q}})\\ &=
        2 (1 - \mathbf{q}\cdot \tilde{\mathbf{q}})
    \end{aligned}
\end{equation}
Similarly, we can modify the above measure to disambiguate the quaternions representations as follows:
\begin{equation}\label{eq:quat_loss_3}
    d_{3}(\mathbf{q},\tilde{\mathbf{q}}) = \min \left\{\left\|\mathbf{q}-\tilde{\mathbf{q}}\right\|^{2},\left\|\mathbf{q}+\tilde{\mathbf{q}}\right\|^{2}\right\}
\end{equation}
All distance measures $d_1$, $d_2$, and $d_3$ address the double-coverage problem, however, the last two are approximations and do not measure the exact geodesic distance.

Similarly, for the scenarios where the joint angles are represented by rotation matrices, we can use the Geodesic distance between a pair of rotation matrices using logarithm map in $SO(3)$ as follows
\begin{equation}\label{eq:rotmat_loss}
    d(\mathbf{R},\tilde{\mathbf{R}}) = \|\log(\tilde{\mathbf{R}}\mathbf{R}^{\top})\|,
\end{equation}
where $\|\log(\tilde{\mathbf{R}}\mathbf{R}^{\top})\|$ is a skew-symmetric matrix containing the rotation axis-angle components and therefore $\|\log(\mathbf{R})\|$ is the magnitude of the angular distance multiplied by a constant. 

\textbf{Hierarchical Loss}:
Proposed approaches in human motion modelling represent each human pose either by 3D joint locations in a global or body’s local coordinate system, or 3D joint angles where, given the limbs' length, the final position and orientation of the body parts are calculated by forward kinematics. Models which use 3D joints locations usually normalize the skeleton size of the training samples and define the loss as an $l_p$ norm over joint locations ~\cite{holden2016,holden2017}. The main problem in such approaches is that during training and generation they are not exploiting the constraints imposed by parameterized skeleton and limbs rigidity. Therefore, the generation phase should be followed by a corrective re-projection onto a valid character skeleton.

Modelling poses by joint angles inherently follows the constraints imposed by parameterized skeleton \cite{taylor2007,martinez2017,fragkiadaki2015,jain2016}. However, defining loss over joint angles ignore the amount of influence that each joint contributes to the learning process and gives all joints equal weights. On the other hand, an error in a parent joint has more impact on the final pose than the same amount of error in its child joints. This is  due to the fact that an error in the parent joints propagates through all of its children down to the leaf nodes in the kinematic tree during forward kinematics. Recently, \cite{pavllo2019} proposed using joint angles to represent body pose but defined the loss over joint locations by applying a differentiable forward kinematics on ground-truth and predicted joint angles. However, applying forward kinematics at each pose is computationally expensive especially for long sequences and when the number of joints is high.

In this work, we propose a hierarchical loss over joint angles which weights each joint's error based on its impact on the reconstructed pose as follows
\begin{equation}\label{eq:hierarchical_loss}
    \mathcal{L}_{rec}(t) = \sum_{k=1}^{K}\alpha_{k}d(\mathbf{X}_{t}^{k}, \tilde{\mathbf{X}}_{t}^{k}),
\end{equation}
where $\mathbf{X}_{t}^{k}$ and $\tilde{\mathbf{X}}_{t}^{k}$ are the ground-truth and the reconstructed joint angles for joint $k$ at time $t$ and $d(.)$ is one of the distance functions defined in Eq.\ref{eq:quat_loss_1}, \ref{eq:quat_loss_2}, \ref{eq:quat_loss_3}, or \ref{eq:rotmat_loss}. $\alpha_{k}$ is the impact factor which weights the impact of the corresponding joint angle on the pose reconstruction. 
A rule of thumb for choosing $\alpha_{k}$s is that the child joint should be weighted with a lower impact factor compared to its parent joint $(\alpha_{k}<\alpha_{parent(k)})$ in the kinematic tree. In this work, we set $\alpha_k$ as the maximum path length from joint $k$ down to all of the connected end-effectors in an average body skeleton. We can define $\alpha_k$ recursively as follows
\begin{equation}
    \alpha_k = \max_{j}(\alpha_j + l_{k-j}), j \in SC_k,
\end{equation}
where $SC_k$ is the set of all children of joint $k$, and $l_{k-j}$ is the length of the bone connecting joints $l$ and $j$. As suggested by \cite{pavllo2019} we also evaluated the results by applying forward kinematics and computed the positional loss.
In practice, the results were very close, while the latter took around $35\%$ longer for training.
%---------------------------------------------------
\subsubsection{Classifiers Loss}\label{sec:Classifiers Loss}
The classifiers are trained to infer the motion attributes and incorporate a complementary loss from the output of intermediate and final layers. This complementary loss can be defined as
\begin{equation}
        \mathcal{L}_{CL}(t) = \sum_{c=1}^{C}\sum_{l \in L_c}\beta_{(c,l)}d(l(\mathbf{X}_t),l( \tilde{\mathbf{X}_t})),
\end{equation}
where $C$ is the number of classifiers and $L_c$ is the set of layers in $c$th classifier. $d(l(\mathbf{X}),l( \tilde{\mathbf{X}}))$ computes the loss for the output of layer $l$ given ground-truth and reconstructed samples. 
$\beta_{(c,l)}$ is a predefined weight assigned for each layer. We compute the loss by using the $l_2$ norm for intermediate layers and cross-entropy loss for the attribute labels. Details of the classifiers' architecture are given in Table \ref{tab:architecture}.
%-----------------------------------------------------------------------------------
\subsubsection{Intrinsic Rotation Representation Constraints}\label{section:Intrinsic Rotation Representation Constraints}
In order to encourage the model to produce valid rotations, we add some constraint terms to the final objective function based on the representation we use for the joint angles. This helps to better ensure convergence at the beginning of the training process and smooths the optimization landscape. Although we normalize the output of $f_{\text{dec}}^{\mathbf{w}}$, better performance is achieved when these outputs are very close to valid rotations leaving the role of normalizers as only a final correction on very small errors.

For rotation matrices we define two constraints: orthogonality and unit determinant, formulated as follows:
\begin{equation}
\begin{aligned}
    \mathcal{L}_{\text{ang}}(t) &= c_{1}\mathcal{L}_{\text{orth}}(t) + c_{2}\mathcal{L}_{\text{det}}(t) \\
    &=\sum_{k=1}^{K}\left(c_{1}\|\tilde{R}_{t}^{k} (\tilde{R}_{t}^{k})^{\top}-I\|_{2}^{2} + c_{2}|\operatorname{det}(\tilde{R}_{t}^{k})-1|\right)
    \end{aligned}
\end{equation}
where the first term encourages the orthogonality of the output matrices and the second term enforces the model to produce matrices with a unit determinant. We also added Sigmoid activation to the output of $f_{\text{dec}}^{\mathbf{w}}$ to ensure that the elements of the output matrices are in the range of $[0,1]$.

For quaternions we only need to set the unit length constraint %on the versors
\begin{equation}
    \mathcal{L}_{\text{ang}} = \mathcal{L}_{q-norm}
    =\sum_{k=1}^{K}|\|\tilde{\mathbf{q}}_{t}^{k} \|_{2}^{2}-1|
\end{equation}
For axis-angle rotation representation we did not set any constraint as they represent the three degrees of freedom by only three scalars.
%---------------------------------------------------
\section{Implementation and Training}\label{sec:Implementation and Training}
\begin{table}
\centering
\begin{tabular}{p{1.7cm}p{5.3cm}} \\ \toprule
Function & Architecture\\ \midrule\midrule
$f_w$, $f_z$  & $2\times[\text{FC}(128)+\text{ELU}]+ \text{FC}(96)+\text{ELU}$\\ %\midrule
$f_{dec}$  & $2\times[\text{FC}(128)+\text{ELU}]+ \text{FC}(96)+\text{ELU}$\\ %\midrule
$f_{enc}^{\mu}$, $f_{prior}^{\mu}$ & $4\times[\text{FC}(128)+\text{ELU}] + \text{FC}(96)$\\ 
%\midrule
$f_{enc}^{\sigma}$, $f_{prior}^{\sigma}$ & $4\times[\text{FC}(128)+\text{ELU}] + \text{FC}(96) + \text{Softplus}$\\ 
%\midrule
$f_{h}$ & $2\times \text{GRUCell}(512)$\\ 
\midrule%\midrule
$f_{\text{enc}}^{\mathbf{w}}$ & $2\times[\text{FC}(128)+\text{ELU}]+ \text{FC}(96)+\text{ELU}$\\ 
%\midrule
$f_{\text{dec}}^{\mathbf{w}}$ & $2\times[\text{FC}(128)+\text{ELU}]+ \text{FC}(3 \times D_P)$\\ \midrule%\midrule
$\text{Classifiers}$ & $3\times[\text{Conv1D}+\text{ReLU}] + \text{AdaptiveAvgPool1D} +  \text{FC}(N_{C})$\\ 
\midrule%\midrule
Denoising Autoencoder & $2\times[\text{Conv1D}+ \text{ReLU}]+ \text{ConvTranspose1D} + \text{ReLU} + \text{ConvTranspose1D}$\\ 
\bottomrule
\end{tabular}
\caption{The architecture of model components. {\normalfont\ FC(n)} is the abbreviation for Fully Connected linear layer with{\normalfont\ n} nodes. {\normalfont\ Conv1D} and {\normalfont\ ConvTranspose1D} are one-dimensional convolution and transposed convolution layers. {\normalfont\ AdaptiveAvgPool1D} is one-dimensional adaptive average pooling layer.\\}
\label{tab:architecture}
\vspace{-8mm}
\end{table}
%---------------------------------------------------
\subsection{Dataset}\label{sec:Dataset}
We trained and evaluated our model on a subset of AMASS \cite{mahmood2019}, a very large database of human motion which unifies different marker-based motion capture datasets by representing them in a common framework. The kinematic tree is represented by 21 joints and the root (pelvis). We used the MoVi \cite{ghorbani2020} and RuB \cite{troje2002} datasets from AMASS for training and evaluating the main module and the rest of the AMASS data for training the denoising autoencoder.

The control parameters in our model are action type and gender. We used a subset of actions: walking, jogging, jumping, and lifting. The data were split into 150, 25, and 25 subjects for the purpose of training, validation, and testing, respectively. All splits contained male and female subjects in equal proportion. 
%The body shape data was defined by using the first $10$ weights of the shape blend shapes for each subject. Finally, the input to each cell was formed by concatenating the pose and body representation resulting in a $79$ dimensional vector.
%---------------------------------------------------
\subsection{Training Process}\label{sec:Training Process}
The details of the model architecture are given in Table \ref{tab:architecture}. All model components were implemented using the PyTorch library.

For training the hierarchical model (Motion Cell, $f_{\text{enc}}^{\mathbf{w}}$, and $f_{\text{dec}}^{\mathbf{w}}$), we optimized the objective function in Eq.\ref{eq:objective_function} with the joint angle distances computed by Eq .\ref{eq:quat_loss_1} using Adam optimizer \cite{kingma2014} with a learning rate of $0.001$, no weight decay, and a batch size of 64. We also set the gradient norm clipping to $0.1$ to avoid any exploding gradients. All weights of the model were initialized using Kaiming initialization \cite{he2015}. We trained our network for 1600 epochs which took around 2 hours on a GeForce RTX 2080 Ti GPU. The scheduling of different loss component coefficients during training can be found in the supplementary material.

% The scheduling of different loss component coefficients during training are illustrated in Fig \ref{fig:schedules}.

For each combination of attributes, the initial internal states of the GRU cells is learned as a Gaussian distribution. Then each sequence is initialized by sampling the initial state from the distribution which corresponds to the required attribute. 

We train our recurrent model in a teacher forcing scheme  (i.e. the ground-truth input is provided to the Motion Cell at each time-step during training). Although this is an effective and fast approach for training, the model is prone to exposure bias and risks overfitting to the training data. To address this problem we experimented with three different mitigating strategies: (i) progressively corrupting input by adding Gaussian noise \cite{fragkiadaki2015,jain2016}, (ii) progressively dropping motion words and exposing the model to its own previous output \cite{martinez2017, pavllo2019}, and (iii) adding joint-wise dropout on the input poses \cite{ghosh2017}. We achieved the best results when we used the second strategy with a scheduled drop rate which helps with the problem of foot skating as well.

We trained all classifiers with similar architecture (Table \ref{tab:architecture}) on $50\%$ of training data using Adam optimizer with a learning rate of 0.005 for $30$ epochs. We used Adaptive Average Pooling before the last fully connected layer to adapt the classifier models to different input lengths.

We trained the denoising autoencoder separately and on the rest of the AMASS data. During training $10\%$ of the input dimensions were chosen randomly and corrupted by Gaussian noise with zero mean and standard deviation of $0.5$. We trained this model for $300$ epochs and using the Adam optimizer with a learning rate of $1e-4$ with an exponential decay of  $0.99$ per $10$ epoch.
%-----------------------------------------------------------
\section{Experiments and Evaluation}\label{sec:Experiments and Evaluation}
\subsection{Models and Ablations}\label{sec:Models and Ablations}
For the purpose of comparison, we compared our model with Pavllo et al.'s Quaternet \cite{pavllo2019} and Fragkiadaki et al.'s Encoder-Recurrent-Decoder (ERD) model \cite{fragkiadaki2015}. These two models were trained on the same training data with the same training hyper-parameter optimization techniques as our model. The initial internal state was learned in the same way to our model. Sampling the initial state is the only source of stochasticity in these two models. 

For all models, we used a common generation scheme. Each walking or jogging sequence was generated with $140$ frames, and the first $20$ frames were discarded during evaluation (resulting in $4$ seconds of motion). For the non-periodic actions (jumping and lifting), we terminated the generated sequence when they collapse to the mean pose. 

In order to evaluate the influence of model components, we trained three additional ablated configurations of the our model. In the first ablated configuration, "Proposed(SL)" (for "Single Layer"), we removed the hierarchical encoder $f_{\text{enc}}^{\mathbf{w}}$ and decoder $f_{\text{dec}}^{\mathbf{w}}$, and fed the individual poses directly to the recurrent model. For the second ablation configuration, "Proposed(NL)" (for "Normal Loss"), we disabled the influence of hierarchical loss by setting all $\alpha_k$ coefficients in the Eq. \ref{eq:hierarchical_loss} to $1$. The last ablation, "Proposed(NC)" (for "No Classifier"), disabled the influence of classifiers on the final loss by setting $\lambda_{\text{CL}}$ to zero.

For all evaluated models, we achieved comparable quantitative results between quaternion and rotation matrix representations, both of which outperformed axis-angle representations. Therefore, for the rest of the paper, we only report the results for quaternions.
%---------------------------------------------------------------------
\subsection{Quantitative Evaluation}\label{sec:Quantitative Evaluation}
In this work, we evaluate models based on two main criteria: quality and diversity.  We expect the generated samples to be realistic and coherent with the attributes which are set as control parameters (quality). In addition, we expect the model to generate motions with high diversity and natural stochasticity while still following the manifold of realistic motions (diversity). To codify both criteria in our quantitative evaluation we use the \emph{Inception Score (IS)} \cite{salimans2016} and \emph{Fréchet Inception Distance Score (FID)} \cite{heusel2017} metrics which were originally proposed for image generative models. Both evaluation metrics have been shown to correlate well with human evaluation on generated images.

IS is formulated based on two criteria, diversity and quality, defined as follows:
\begin{equation}\label{eq:inception_score}
    \operatorname{IS}=\exp \left(\mathbf{E}_{\tilde{\boldsymbol{X}} \sim p_{g}} D_{K L}(p(\mathbf{a} | \tilde{\mathbf{X}}) \| p(\mathbf{a}))\right)
\end{equation}
where $\mathbf{\tilde{X}}$ is a synthetic sample generated by a generative model, $p(\mathbf{a} | \tilde{\mathbf{X}})$ is the conditional attribute distribution of a classifier which is pre-trained on separate training data, and $p(\mathbf{a})=\int_{\tilde{\mathbf{X}}} p(\mathbf{a} | \tilde{\mathbf{X}}) p_{\text{model}}(\tilde{\mathbf{X}})$ is the marginal attribute distribution. Equation \ref{eq:inception_score} can be also written as $
    \operatorname{IS}=\exp \left(H(\mathbf{a})-H(\mathbf{a} | \mathbf{\tilde{X}})\right)
$
, where $H(\mathbf{a})$ and $H(\mathbf{a} | \tilde{\mathbf{X}})$ are the attribute entropy and the conditional attribute entropy, respectively. Generated animations which fulfil the semantics defined by the attributes should have a conditional attribute distribution $p(\mathbf{a} | \tilde{\mathbf{X}})$ with low entropy. In other words, the classifier should be very confident about the attribute associated with the generated animation. On the other hand, we expect our model to generate a high variety of motions for each attribute class, therefore, $p(\mathbf{a})$ should have a high entropy. An estimator of IS as follows
\begin{equation}
    \operatorname{IS} \approx \exp{\left(\frac{1}{M} \sum_{i=1}^{M} D_{K L}\left(p(\mathbf{a} | \tilde{\mathbf{X}}^{(i)} \| \hat{p}(\mathbf{a}))\right)\right)},
\end{equation}
where $\tilde{\mathbf{X}}^{(i)}$ is a generated motion sample and $\hat{p}(\mathbf{a})=\frac{1}{M} \sum_{i=1}^{M} p\left(\mathbf{a}| \tilde{\mathbf{X}}^{(i)}\right)$ is the empirical conditional distribution.

FID captures the similarity between the generated and the real motion samples. It evaluates the model by comparing the statistics of a set of generated samples to a set of real motion sequences from the dataset. Similar to IS, we use a classifier trained on a separate dataset. Then, the activations of the last feature extraction layer (the last layer prior to the last fully connected layer) are summarized as a multivariate Gaussian distribution for synthetic and real data. The distance between the two distributions is then computed with Fréchet Distance as follows:
\begin{equation}
    \operatorname{FID}=\left\|\mu_{g}-\mu_{d}\right\|_{2}^{2}+\operatorname{Tr}\left(\Sigma_{g}+\Sigma_{d}-2\left(\Sigma_{g} \Sigma_{d}\right)^{\frac{1}{2}}\right),
\end{equation}
where $\mathcal{N}\left(\mu_{g}, \Sigma_{g}\right)$ and $\mathcal{N}\left(\mu_{d}, \Sigma_{d}\right)$ are the distributions of the activations in the last feature extraction layer for synthetic and real data, respectively.

\begin{table}
\centering
\begin{tabular}{p{3cm}p{.7cm}p{1cm}} \\ \toprule
Model & IS $\uparrow$ & FID $\downarrow$\\ \midrule\midrule
Quaternet \cite{pavllo2019} & 5.12 & 92.31\\
% \midrule
ERD \cite{fragkiadaki2015} & 5.91 & 86.42 \\
% \midrule
Proposed(SL) & 6.45 & 31.41\\
% \midrule
Proposed(NL) & 7.11 & 17.3\\
% \midrule
Proposed(NC) & 7.43  & 11.92\\
% \midrule
Proposed & 7.52 & 10.45\\
\midrule
% \midrule
Real data & 7.64 & 0\\
\bottomrule
\end{tabular}
\caption{Results from quantitative evaluations using IS (higher score is better) and FID (lower score is better).}
\label{tab:quantitative_results}
%\vspace{-5mm}
\end{table}

\textbf{Results}: The results of quantitative evaluations are shown in Table \ref{tab:quantitative_results}. Using each model, we generated $1000$ samples for each combination of attributes ($M=8000$ in total). Quaternet and ERD generated convincing walking and jogging samples. However, since the only source of stochasticity is the initial hidden state, these models fail to generate a diverse set of sequences (ERD's performance was slightly better due to its hierarchical structure yielding higher diversity at the beginning of the motions). In addition, they usually failed to generate a complete sequence for non-periodic motion such as lifting and were regressed to the mean pose after $70-80$ frames. Among ablation configurations, Proposed(SL) had the lowest performance showing the significant influence of hierarchical structure in generating diverse motions (higher $H(\mathbf{a})$). The lower scores for the other two ablation configurations (Proposed(NL) and Proposed(NC)) indicate the impact of our hierarchical loss structure and the effect of integrating classifiers into the loss, respectively, on the conditional attribute entropy and higher motion quality.

\subsection{Qualitative Evaluation}\label{sec:Qualitative_Evaluation}
\begin{table*}
\centering
\begin{tabular}{p{3cm}p{2cm}p{2cm}p{2cm}p{2cm}p{2cm}} \\ \toprule
\multicolumn{6}{c}{Actions} \\
\cmidrule(r){2-5}
Model & Walking & Jogging & Jumping & Lifting & Average\\ \midrule\midrule
Quaternet \cite{pavllo2019} & $9.31{\displaystyle \pm }0.25$ & $9.45{\displaystyle \pm }0.22$ & $6.27{\displaystyle \pm }0.31$ & $5.87{\displaystyle \pm }0.28$ & $7.73{\displaystyle \pm }0.26$\\
% \midrule
ERD \cite{fragkiadaki2015} & $8.97{\displaystyle \pm }0.33$ & $8.32{\displaystyle \pm }0.30$ & $6.31{\displaystyle \pm }0.28$ & $5.71{\displaystyle \pm }0.23$ & $7.31{\displaystyle \pm }0.28$\\
% \midrule
Proposed(SL) & $9.15{\displaystyle \pm }0.24$ & $9.26{\displaystyle \pm }0.22$ & $8.43{\displaystyle \pm }0.37$ & $8.62{\displaystyle \pm }0.18$ & $8.87{\displaystyle \pm }0.25$\\
% \midrule
Proposed(NL) & $9.03{\displaystyle \pm }0.29$ & $8.97{\displaystyle \pm }0.21$ & $8.25{\displaystyle \pm }0.19$ & $8.30{\displaystyle \pm }0.32$ & $8.64{\displaystyle \pm }0.26$\\
% \midrule
Proposed(NC) & $9.32{\displaystyle \pm }0.27$ & $9.42{\displaystyle \pm }0.30$ & $8.95{\displaystyle \pm }0.31$ & $8.92{\displaystyle \pm }0.34$ & $9.15{\displaystyle \pm }0.27$\\
% \midrule
Proposed & $9.38{\displaystyle \pm }0.13$ & $9.35{\displaystyle \pm }0.12$ & $9.13{\displaystyle \pm }0.28$ & $9.27{\displaystyle \pm }0.27$ & $9.28{\displaystyle \pm }0.21$\\
% \midrule
\midrule
Real data & $9.64{\displaystyle \pm }0.13$ & $9.81{\displaystyle \pm }0.24$ & $9.62{\displaystyle \pm }0.1$ & $9.53{\displaystyle \pm }0.21$ & $9.65{\displaystyle \pm }0.17$\\
\bottomrule
\end{tabular}
\caption{Results for qualitative evaluation. The results show the average scores assigned to a set of motions sampled from each action set and from each model, where 1 corresponds to completely unrealistic and 10 corresponds to completely realistic.}
\label{tab:qualitative_results}
\vspace{-5mm}
\end{table*}

\begin{figure}[t]
  \centering
  \begin{subfigure}[b]{.23\textwidth}
  \includegraphics[width=1\linewidth]{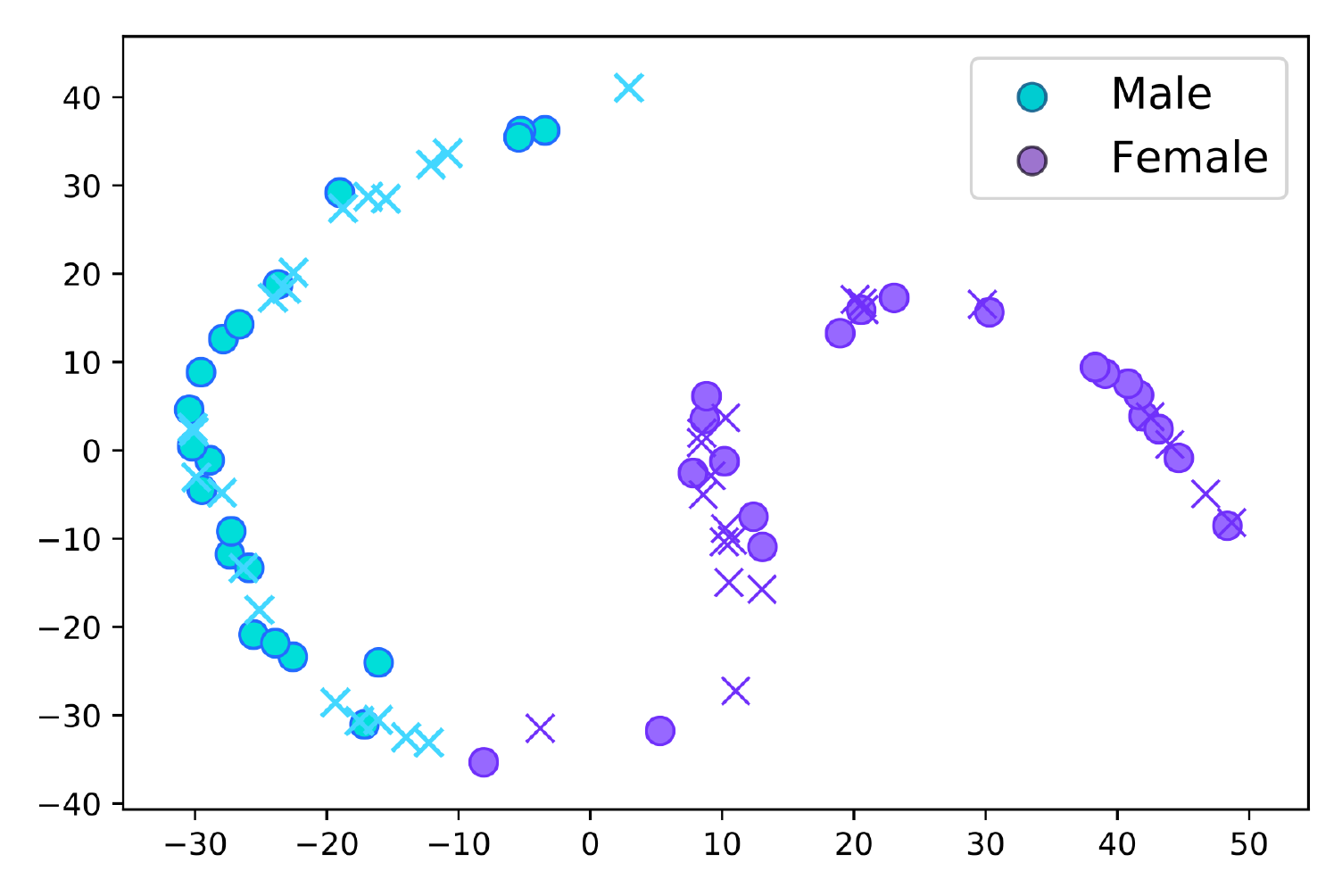}
  \caption{Samples for different genders}
\end{subfigure}
  \begin{subfigure}[b]{.23\textwidth}
  \includegraphics[width=1\linewidth]{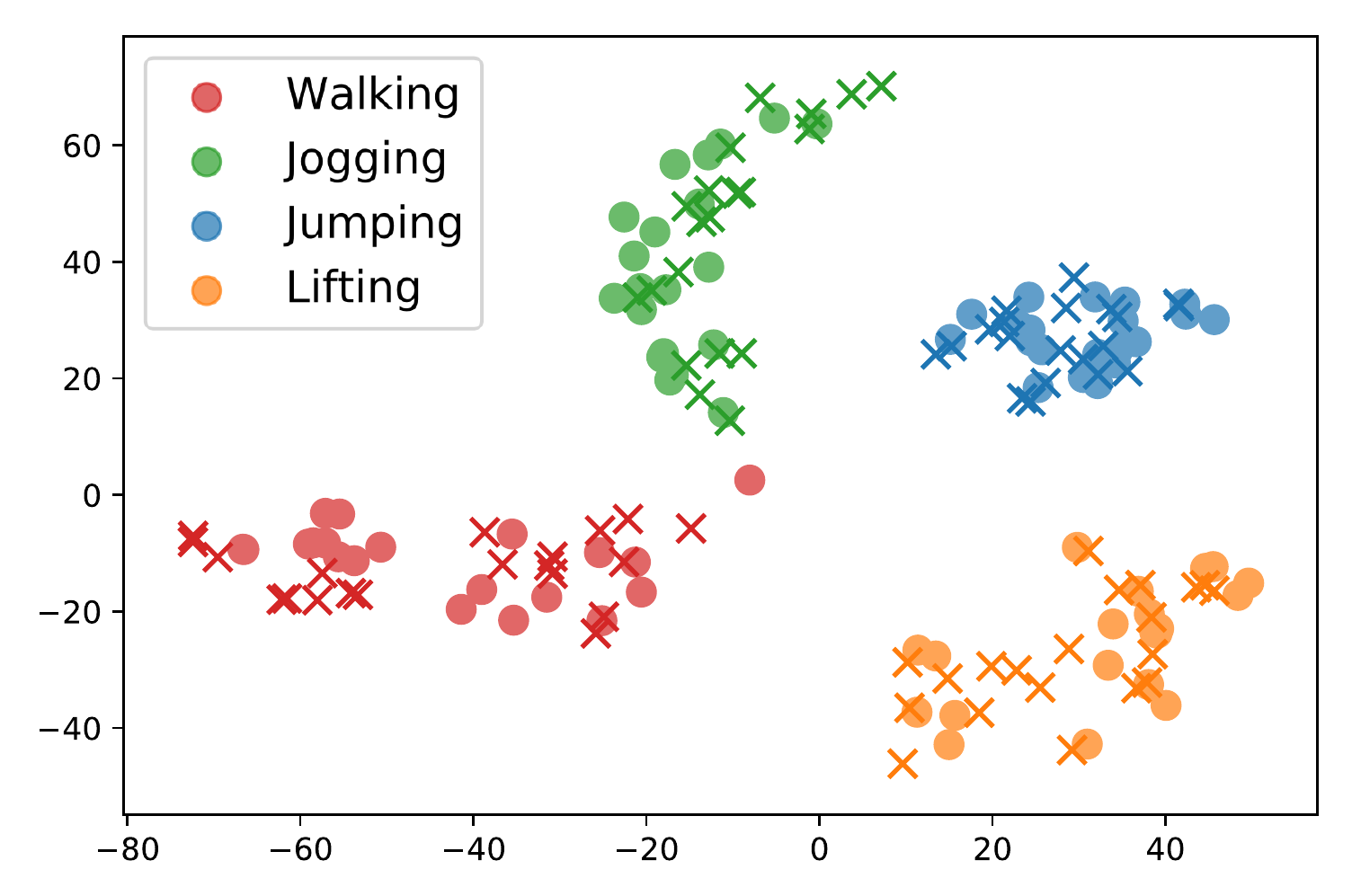}
  \caption{Samples for different actions}
\end{subfigure}
  \caption{\label{fig:tsne}
           Visualizing the activations of the last layer of action classifier projected onto two dimensions using t-sne for real data (circles) and data synthesized data (crosses) by our model. As can be seen, synthesized samples strongly coincide with the corresponding clusters formed by real data.}
   \vspace{-5mm}
\end{figure}
To qualitatively evaluate how realistic and natural the synthetic animations are we also performed an experiment for subjective evaluation. All six evaluated models in the previous section were used in this experiment as well. We sampled five sequences for each of the four action types from the synthesized sequences of each model resulting in $6 * 4 * 5 = 120$ synthesized samples which were added to $20$ motion sequences from real data. $20$ human observers rated each motion sample from $1$ (completely unrealistic) to $10$ (completely realistic). The motion clips were displayed to the raters in a randomized order. Raters were asked to rate each animation after it was displayed completely, and no information about the aim of the experiment was given to the raters. We used a few motion clips for the purpose of training the raters before each experiment. 

\textbf{Results: }The results of our qualitative evaluation are illustrated in Table \ref{tab:qualitative_results}. The qualitative results correlate well with the quantitative results (Pearson correlations of $0.86$ and $-0.95$ with IS and FID, respectively). Quaternet and ERD achieved the lowest ratings for non-periodic actions (jumping and lifting) since they usually fail to complete these motions and instead regressed to the mean pose in the last frames. Among ablation configurations, Proposed(NL) achieved the lowest mean rating, which shows the impact of hierarchical loss on having more realistic motions. Although the main goal of our hierarchical structure is to improve the diversity of motions, the lower ratings achieved by Proposed(SL) compared to the main model nevertheless demonstrates the effectiveness of our hierarchical architecture even on short sequences. Similar to the quantitative evaluation, disabling the hierarchical structure of the loss function (Proposed(NL)) resulted in decreased ratings, suggesting the importance of classifiers to better learn action modes on the motion manifold.

Figure \ref{fig:tsne} shows the visualization of motion sequences in a two-dimensional space. We sampled $20$ sequence for each action type and gender from real data (circles) and the sequences generated by our model (crosses), extracting the activations from the last layer of classifiers. We then applied \emph{t-sne} \cite{maaten2008} dimensionality reduction to project the activations onto two dimensions. As seen, our model generates sequences with similar diversity to real data while still accurately separating the modes for each action type and gender.
%-------------------------------------------------------------------------
\subsection{Discussion}\label{sec:Discussion}
In this work, we propose a motion generative model with a focus on preserving the stochastic nature of human motion while generating convincing and natural spatiotemporal motion sequences. The proposed model uses a deep hierarchical recurrent framework which can further be tuned via weak control signals such as action type. Each sequence is generated using a probabilistic recurrent structure which models the underlying stochasticity by injecting noise in an abstract level. We also propose a novel hierarchical geodesic loss which incorporates the structural information of the kinematic tree and compares joint angles based on angular distances, yielding a better representation of error and more accurate learning.

\emph{Limitations and Future Work.}
The proposed architecture was implemented for four different action types in addition to the gender attribute. Extending the model to include more actions is not straightforward and is prone to mean collapse. Different strategies can possibly be exploited to increase the capacity of the architecture for more action types or other additional semantics. One possible solution is to increase the network capacity, though we expect this may make training more difficult. Another solution could be to train a separate network for each subset of attributes or actions, which may serve as a feasible solution since our network is relatively small (around 30MB). Finally, providing additional strong control signals such as body contact with the environment could serve to decrease the uncertainty in the motion generation phase and prevent it from collapsing to the mean pose. These control signals could be provided manually by the animator or by a separate network which is trained on the data~\cite{pavllo2019, holden2017}. 

In our recurrent model, each cell is represented as a VAE conditioned on the previous internal state. However, VAEs are based on maximizing a log-likelihood lower bound which might give a suboptimal solution for the true log-likelihood. We also made a strong assumption about the posterior distribution by modelling it as a standard isotropic Gaussian, which could increase the error in the posterior approximation. It is possible that normalizing flows~\cite{kobyzev2019,papamakarios2019} could be exploited to improve this aspect of the model. The approximate posterior distribution could be parameterized by a network of normalizing flows which is applied to the output of the VAE decoder; this has been shown to provide a tighter lower bound for other applications \cite{rezende2015}. Alternatively, the VAE module in our Motion Cell could potentially be replaced entirely with a conditional normalizing flow network in which the temporal dependencies are modelled by conditioning the prior on the previous internal state.
%-------------------------------------------------------------------------
\FloatBarrier
%\subsection{References}
\bibliographystyle{eg-alpha-doi} 
\bibliography{egbibsample}       

% biblatex with biber
% \printbibliography  

%-------------------------------------------------------------------------
\clearpage
\appendix
\setcounter{figure}{0}
\section{Supplementary Material}\label{sec:Supplementary Material}
%-------------------------------------------------------------------------
\subsubsection{Recurrent Neural Networks}\label{sec:recurrent_neural_network}
Most of proposed methods \cite{pavllo2019, jain2016, fragkiadaki2015, martinez2017} for modelling human motion are based on recurrent neural networks (RNN). An RNN \cite{cho2014,hochreiter1997} models data recursively by decomposing the probability distribution of the sequence over time
\begin{equation}\label{eq:autoregressive}
    P_{\theta}\left(\mathbf{x}_{1}, \ldots, \mathbf{x}_{T}\right)=\prod_{t=2}^{T} P_{\theta}\left(\mathbf{x}_{t} | \mathbf{x}_{< t}\right) P_{\theta}\left(\mathbf{x}_{1}\right),
\end{equation}
where $\theta$ is the set of model parameters and $P_{\theta}\left(\mathbf{x}_{1}, \ldots, \mathbf{x}_{T}\right)$ is the likelihood of sequence $\mathbf{x}_{1:T}$. Each time step includes two main operations: updating the hidden internal state of the model which summarizes the past information, and a mapping from the hidden state to the next element in the sequence. Therefore, at each time-step we have
\begin{equation}
    \mathbf{h}_{t} = f_{h}(\mathbf{h}_{t-1}, \mathbf{x}_t)
\end{equation}
\begin{equation}
\begin{aligned}
    P_{\theta}\left(\mathbf{x}_{t+1} | \mathbf{x}_{\leq t}\right) &=
    P_{\theta}\left(\mathbf{x}_{t+1} | \mathbf{h}_{t}\right)\\ &=f_{o}(\mathbf{h}_{t}),
\end{aligned}
\end{equation}
where $\mathbf{h}_{t}$ is the internal hidden state and $f_h$ is the non-linear updating function parameterized by $\theta$. $f_{o}$ is the mapping from internal state to the output and usually characterized by another network. 
%-------------------------------------------------------------------------
\subsubsection{Variational Autoencoder}
The Variational Autoencoder (VAE) \cite{Kingma2013} is a class of deep generative models which optimize a deep latent-variable model (DLVM) jointly with the inference model using a gradient-based optimizer. DLVM is a class of models with a simple prior $p_{\theta}(\mathbf{z})$ and complex marginal $p_{\theta}(\mathbf{x})$ and likelihood $p_{\theta}(\mathbf{x}|\mathbf{z})$ distributions, where $z$ is a latent variable and $x$ is the observed variable. VAEs approximate the intractable posterior inference $p_{\theta}(\mathbf{z}|\mathbf{x})$ by introducing a parametric inference function $q_{\phi}(\mathbf{z}|\mathbf{x})$. Variational parameters $\phi$ and the generative parameters $\theta$ can be optimized jointly by maximizing the \emph{evidence lower bound} (ELBO) defined as follows:
\begin{equation}\label{eq:ELBO}
\begin{aligned}
    \mathcal{E}_{\boldsymbol{\theta}, \boldsymbol{\phi}}(\mathbf{x})&=
     \mathbf{E}_{\mathbf{z} \sim q_{\phi}(\mathbf{z} | \mathbf{x})} \log \frac{p_{\theta}(\mathbf{x} , \mathbf{z})}{q_{\phi}(\mathbf{z} | \mathbf{x})}\\
     &=
    \mathbf{E}_{\mathbf{z} \sim q_{\phi}(\mathbf{z} | \mathbf{x})} \log p_{\theta}(\mathbf{x} | \mathbf{z}) - D_{\mathrm{KL}}\left(q_{\phi}(\mathbf{z} | \mathbf{x}) \| p_{\theta}(\mathbf{z})\right).
\end{aligned}
\end{equation}
The first term is the expected log-likelihood or reconstruction term which is estimated by Monte Carlo estimator and reparameterization trick \cite{Kingma2013}. The second term, $D_{\mathrm{KL}}\left(q_{\phi}(\mathbf{z} | \mathbf{x}) \| p_{\theta}(\mathbf{z})\right)$, is the Kullback-Leibler divergence between $q_{\phi}(\mathbf{z}| \mathbf{x})$ and  $p_{\theta}(\mathbf{z})$ which acts as a specific regularization term \cite{kingma2019}.
The combination of inference $q_{\phi}$ and generative $p_{\theta}$ models form a kind of autoencoder where the first term in Eq.\ref{eq:ELBO} is called the reconstruction term and the second term is called the regularization term. Vanilla VAEs suffer when the space of outputs is multi-modal, resulting in blurry generated samples. Sohn et al. \cite{sohn2015} proposed conditional VAE (CVAE) as an extension of VAE for learning structured output predictions simply by conditioning the generative process on an control variable. CVAE not only helps to address the problem of one-to-many mapping but also allows the model to control the generated sample class and/or characteristics by the control variable. The ELBO for CVAE is formulated as follows:
\begin{equation}\label{eq:CVAE_ELBO}
    \mathcal{E}_{\boldsymbol{\theta}, \boldsymbol{\phi}}(\mathbf{x}, \mathbf{a})=
    \mathbf{E}_{\mathbf{z} \sim q_{\phi}(\mathbf{z} | \mathbf{x}, \mathbf{a})} \log p_{\theta}(\mathbf{x} | \mathbf{z}, \mathbf{a}) - D_{\mathrm{KL}}\left(q_{\phi}(\mathbf{z} | \mathbf{x}, \mathbf{a}) \| p_{\theta}(\mathbf{z}|\mathbf{a})\right),
\end{equation}
where $\mathbf{a}$ is the input variable.
% [CALDEN] - For the sake of symmetry in subsections, it would be good to have a line or two here which comments on the appropriateness of VAEs (or CVAEs) to the problem to match with the closing lines of Section 3.1
\subsubsection{RVAE Objective Function}\label{sec:RVAE Objective Function}
Here we explain how the RVAE objective function in Eq. \ref{eq:vae_loss} is derived.
We can consider the whole sequence as a single sample and write the ELBO similar to Eq. \ref{eq:CVAE_ELBO} as follows:
\begin{equation}\label{eq:R_ELBO}
\begin{aligned}
    \mathcal{E}&=
     \mathbf{E}_{q_{\phi}(\mathbf{z}_{\leq N} | \mathbf{w}_{\leq N}, \mathbf{a}_{\leq N})} \log \frac{p_{\theta}(\mathbf{w}_{\leq N} , \mathbf{z}_{\leq N} |  \mathbf{a}_{\leq N})}{q_{\phi}(\mathbf{z}_{\leq N} |  \mathbf{w}_{\leq N}. \mathbf{a}_{\leq N})}
\end{aligned}
\end{equation}
By factorizing $p_{\theta}$ and $q_{\phi}$ across time (similar to \ref{eq:autoregressive}) we have
\begin{equation}\label{eq:R_ELBO2}
\begin{aligned}
    \mathcal{E}&=
     \mathbf{E}_{q_{\phi}(\mathbf{z}_{\leq N} | \mathbf{w}_{\leq N}, \mathbf{a}_{\leq N})} 
     \log 
     \frac
     {\prod_{n=1}^{N}p_{\theta}(\mathbf{w}_{n} , \mathbf{z}_{n} |  \mathbf{w}_{< n} , \mathbf{z}_{< n}, \mathbf{a}_{\leq n})}
     {\prod_{n=1}^{N}q_{\phi}(\mathbf{z}_{n} |  \mathbf{z}_{< n}\mathbf{w}_{< n}. \mathbf{a}_{\leq n})}\\
     &=
     \mathbf{E}_{q_{\phi}(\mathbf{z}_{\leq N} | \mathbf{w}_{\leq N}, \mathbf{a}_{\leq N})} 
     \log
     \prod_{n=1}^{N}
     \frac
     {p_{\theta}(\mathbf{w}_{n}|  \mathbf{w}_{< n} , \mathbf{z}_{\leq n}, \mathbf{a}_{\leq n})p_{\theta}(\mathbf{z}_{n}|\mathbf{z}_{< n}, \mathbf{a}_{\leq n})}
     {q_{\phi}(\mathbf{z}_{n} |  \mathbf{z}_{< n}\mathbf{w}_{< n}. \mathbf{a}_{\leq n})}\\
     &=
     \mathbf{E}_{q_{\phi}\left(\mathbf{z}_{\leq N} | \mathbf{w}_{\leq N}, \mathbf{a}_{\leq N} \right)}\biggr[\sum_{n=1}^{N}\log p_{\theta}\left(\mathbf{w}_{n} | \mathbf{w}_{<n},  \mathbf{z}_{\leq n}, \mathbf{a}_{\leq n}\right)\\ &- \lambda_{KL}\mathrm{KL}(q_{\phi}(\mathbf{z}_{n} | \mathbf{w}_{\leq n}, \mathbf{z}_{<n},  \mathbf{a}_{\leq n}) \| p_{\theta}(\mathbf{z}_{n} | \mathbf{w}_{<n}, \mathbf{z}_{<n},  \mathbf{a}_{\leq n}))\biggr],
\end{aligned}
\end{equation}
where the RVAE objective is the negative value of the above ELBO.
\subsection{More Training and Synthesis Details}
\subsubsection{Scheduling Loss Coefficients}\label{sec:Scheduling Loss Coefficient}
\begin{figure}[h]
  \centering
  \includegraphics[width=.9\linewidth]{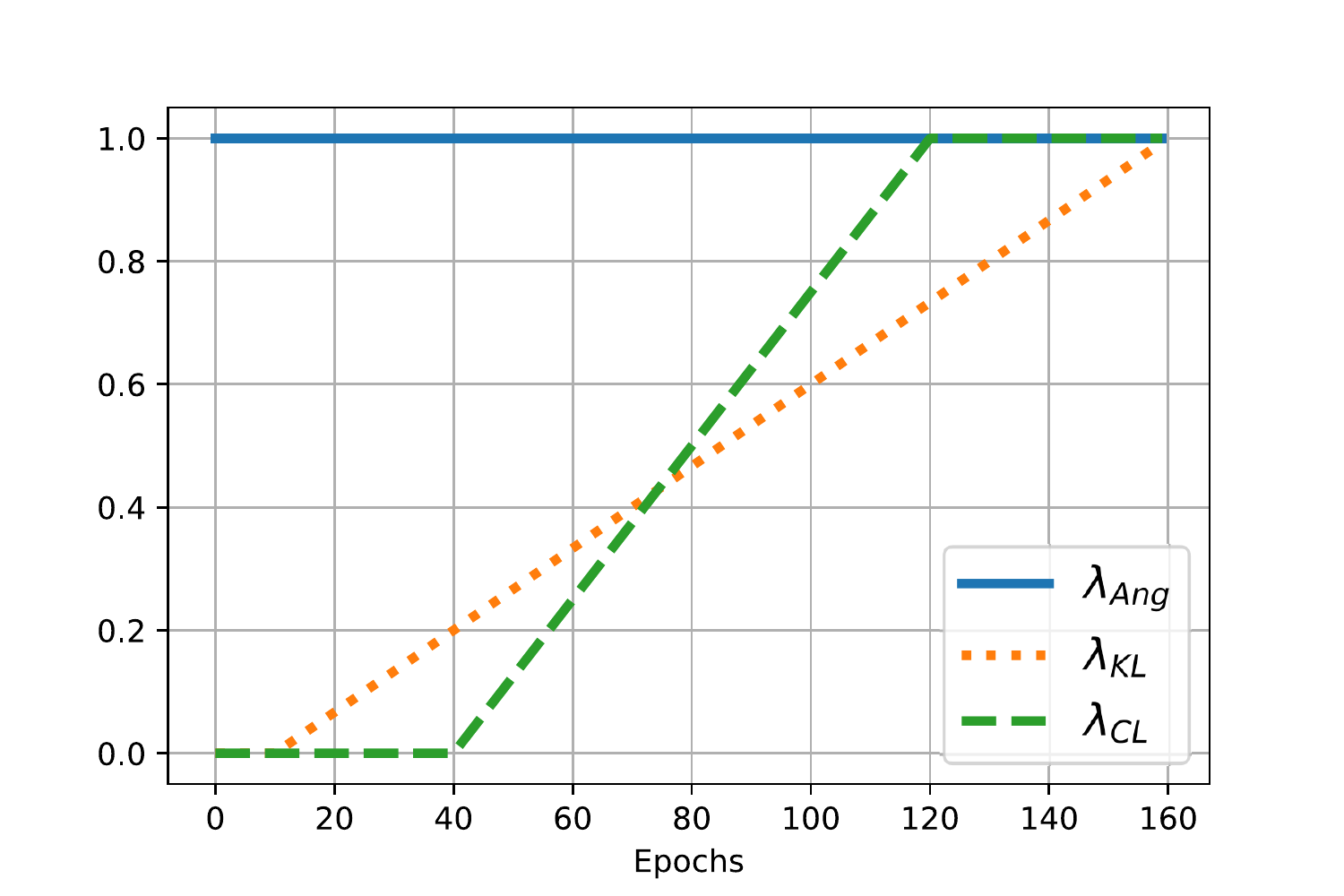}
  \caption{\label{fig:schedules}
           Scheduling the coefficients for each term in Eq. \ref{eq:objective_function} during training. }
           \vspace{-5mm}
\end{figure}

Figure \ref{fig:schedules} shows how we set the schedulers for loss coefficients in Eq. \ref{eq:objective_function}.
We set an annealing scheduler for $\lambda_{KL}$ to address the KL vanishing problem. We also set $\lambda_{CL}$ to zero for the beginning of the training and then gradually increase it. These two strategies allow the model to focus more on capturing useful information for reconstruction during the initial epochs.

\subsubsection{Hardware and Software}
We implemented the framework using PyTorch library. We also used a single GeForce RTX 2080 Ti GPU for parallel computing both in training and synthesis.

\subsubsection{Visualization and Sequence Samples}
The visualization of motion sequences for conducting qualitative experiments was shown as stick figures. For the demo, we used Unity to visualize SMPL \cite{Loper2015} model with average body shape for both females and males.

We present representative samples of the output of our model in Figure 2. You can see by comparing the generated samples (orange) to real samples (blue), our synthesized samples look very natural and from the same action cluster. For more samples please check out our video demo.

To show the ability of our framework in learning transitions between different actions, we also trained the network on ACCAD database \cite{ACCAD} which contains samples of transition between jogging and walking. The examples of generated samples are included in the video demo.

\begin{figure*}[t]
  \centering
  \begin{subfigure}[h]{.8\textwidth}
    \includegraphics[width=\linewidth]{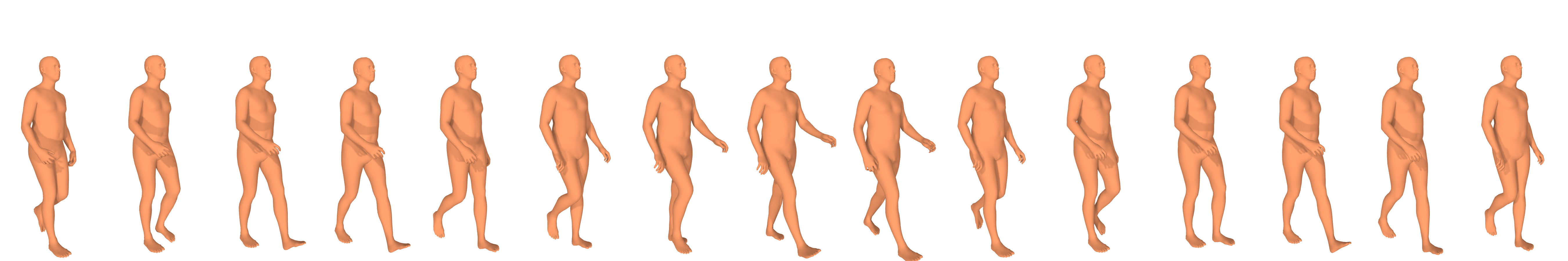}
  \end{subfigure}
  \begin{subfigure}[h]{.8\textwidth}
    \includegraphics[width=\linewidth]{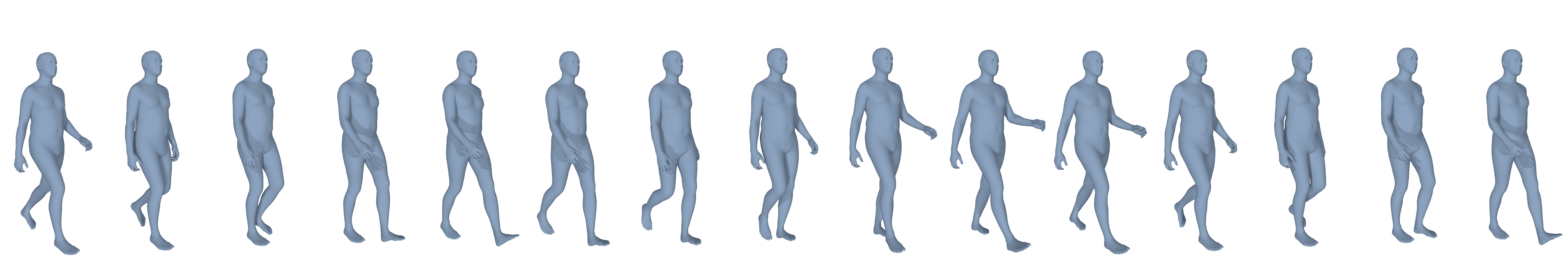}
  \end{subfigure}
  \begin{subfigure}[h]{.8\textwidth}
    \includegraphics[width=\linewidth]{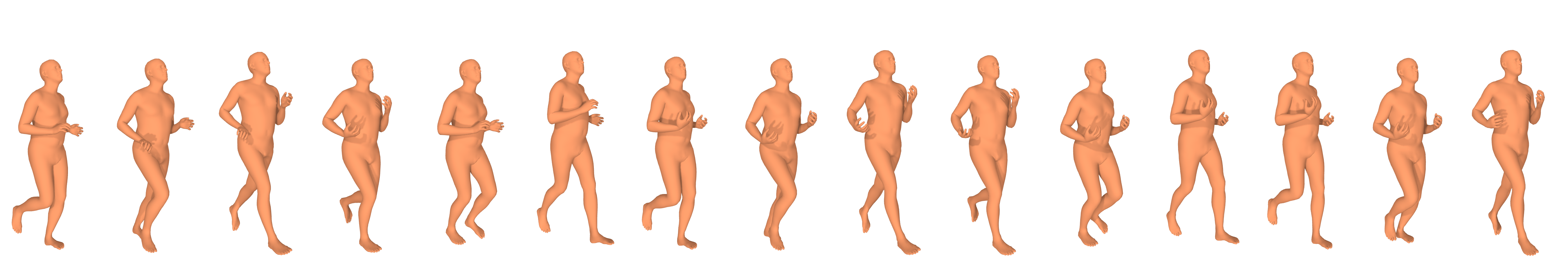}
  \end{subfigure}
  \begin{subfigure}[h]{.8\textwidth}
    \includegraphics[width=\linewidth]{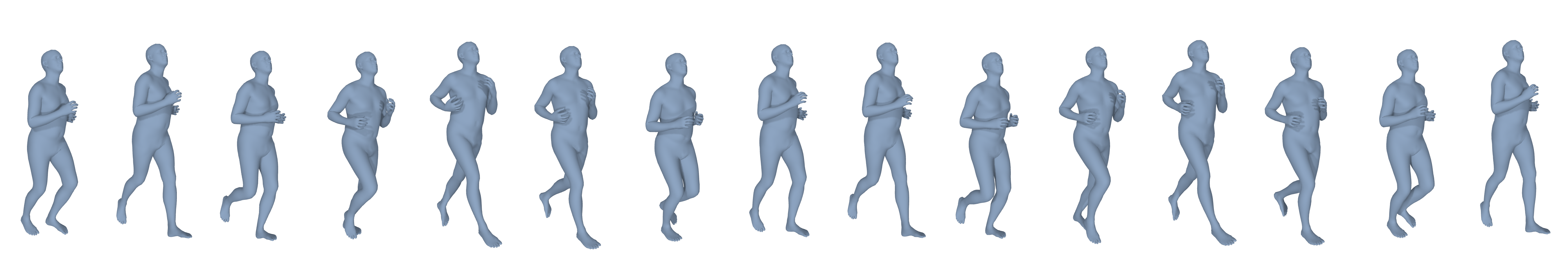}
  \end{subfigure}
    \begin{subfigure}[h]{.8\textwidth}
    \includegraphics[width=\linewidth]{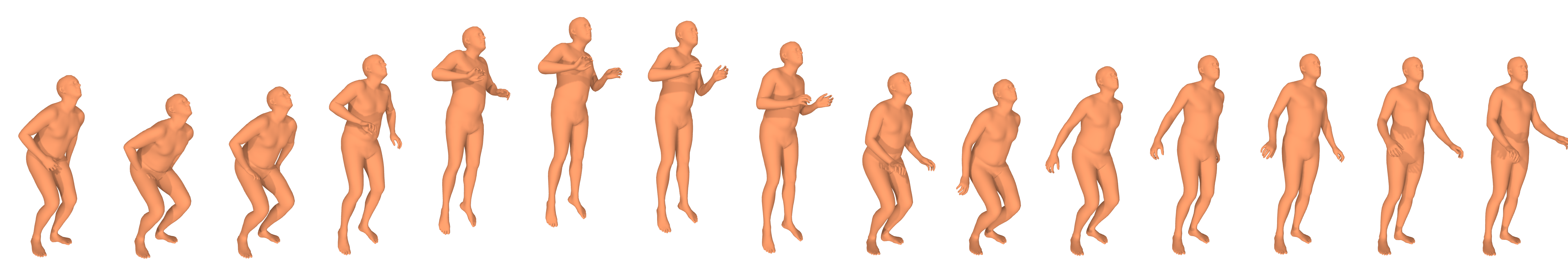}
  \end{subfigure}
  \begin{subfigure}[h]{.8\textwidth}
    \includegraphics[width=\linewidth]{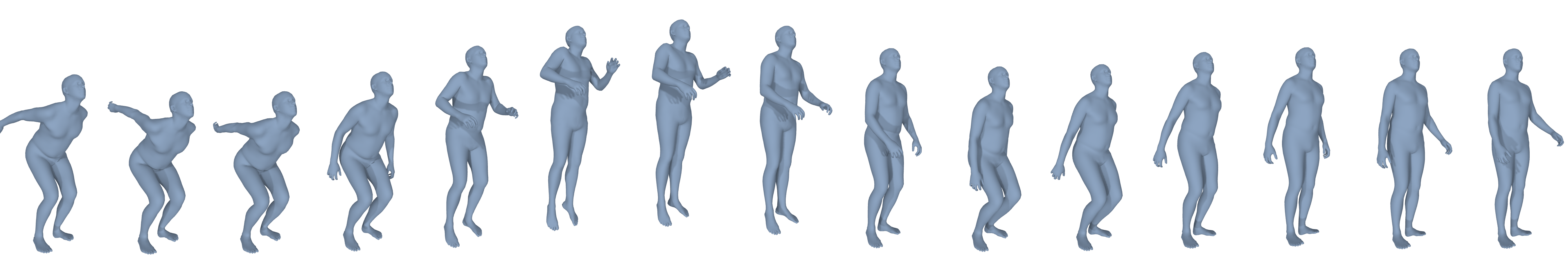}
  \end{subfigure}
    \begin{subfigure}[h]{.8\textwidth}
    \includegraphics[width=\linewidth]{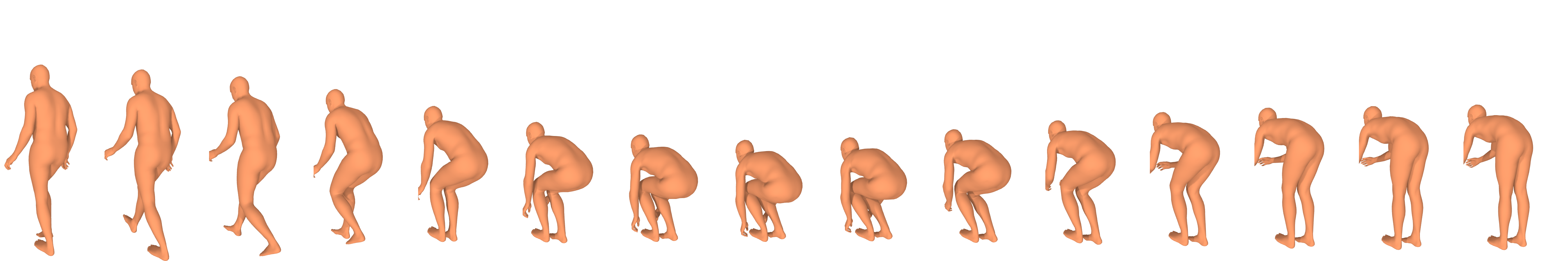}
  \end{subfigure}
  \begin{subfigure}[h]{.8\textwidth}
    \includegraphics[width=\linewidth]{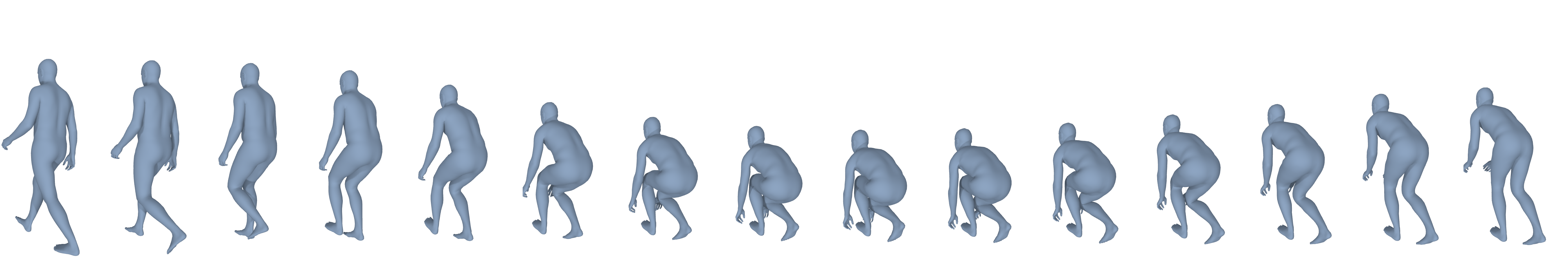}
  \end{subfigure}
  \caption{Samples of real (blue) and synthetic (orange) motion sequences.}
  \label{fig:samples}
\end{figure*}

%-------------------------------------------------------------------------

\end{document}